\documentclass[vecphys]{svmult}

\usepackage{makeidx}         
\usepackage{graphicx}        
\usepackage{multicol}        
\usepackage[bottom]{footmisc}

\makeindex             

\newcommand\lsim{\mathrel{\rlap{\lower4pt\hbox{\hskip1pt$\sim$}}
    \raise1pt\hbox{$<$}}}
\newcommand\gsim{\mathrel{\rlap{\lower4pt\hbox{\hskip1pt$\sim$}}
    \raise1pt\hbox{$>$}}}

\newcommand\vev[1]{{\langle {#1} \rangle}}
\renewcommand\({\left(}
\renewcommand\){\right)}
\renewcommand\[{\left[}
\renewcommand\]{\right]}

\newcommand\eq[1]{Eq.~(\ref{#1})}
\newcommand\eqs[2]{Eqs.~(\ref{#1}) and (\ref{#2})}

\newcommand\eqreff[1]{(\ref{#1})}

\newcommand\pa{\partial}

\newcommand\ee{\end{equation}}
\newcommand\be{\begin{equation}}
\newcommand\eea{\end{eqnarray}}
\newcommand\bea{\begin{eqnarray}}

\newcommand\mpl{M_{\rm P}}

\newcommand{\dlabel}[1]{\label{#1} \ \ \ \ \ \ \ \ #1\ \ \ \ \ \ \ \ }

\def\calp{{\cal P}}

\newcommand\TeV{\,\mbox{TeV}}
\newcommand\GeV{\,\mbox{GeV}}
\newcommand\MeV{\,\mbox{MeV}}

\newcommand\eV{\,\mbox{eV}}

\newcommand\msun{M_\odot}

\newcommand\sub[1]{_{\rm #1}}

\newcommand\mone{^{-1}}
\newcommand\mtwo{^{-2}}

\newcommand\mhalf{^{-1/2}}
\newcommand\half{^{1/2}}

\newcommand\quarter{^{1/4}}

\newcommand\diff{d\,}

\begin{document}

\title*{Particle physics models of inflation}
\author{David H.\ Lyth\inst{1}}
\institute{Physics Department, Lancaster University, Lancaster
LA1 4YB, UK}
%
%
\maketitle

\abstract{
Inflation models are compared with observation on the assumption that the 
curvature perturbation is generated from the vacuum fluctuation of the 
inflaton  field. The focus is on single-field models  with canonical kinetic 
terms, classified as small- medium- and large-field according to the 
variation of the inflaton field while cosmological scales  leave the horizon.
Small-field models are constructed according to the usual paradigm for beyond
Standard Model physics.\footnote{Based on a talk given at the 22nd IAP Colloquium, ``Inflation +25'', Paris, June 2006} 

\section{Introduction}

Several different types of inflation model  have been proposed over the
years. In this survey they are compared  with
observation on the assumption that the curvature perturbation is generated
during inflation. The survey is based on works with my collaborators,
in particular \cite{treview,book,bl,al}.

I focus largely 
on the slow-roll paradigm, because it is the simplest and most
widely-considered possibility.
It assumes that  the energy
density and pressure dominated by the scalar field potential $V$, whose value
hardly varies during one Hubble time. Unless otherwise stated, we consider
single-field inflation, where just one canonically-normalized `inflaton'
field $\phi$  has  significant time-dependence.

In the vacuum, $V=0$. To generate the inflationary value of $V$, one
or more fields must be strongly displaced form the vacuum and there are
two simple possibilities. In {  non-hybrid} inflation,
$V$ is generated almost entirely by the displacement of the inflaton
field from its vacuum, while in { hybrid}
 models  it is generated almost
entirely by the displacement of some other  field $\chi$,
called the waterfall field 
because its eventual descent to the vacuum is supposed to be
very rapid. Hybrid models are not at all artificial, being based on the
concept of spontaneous symmetry breaking and restoration which is ubiquitous
in early-universe cosmology.

The first slow-roll   model,
 termed New Inflation \cite{new} (see also \cite{hm}),
 was non-hybrid. It made contact with  particle physics
through the use of a GUT theory, but  was
quickly seen to generate too big a curvature perturbation
\cite{specpred}.
Viable models using a GUT and supersymmetry were developed,
including   \cite{earlyhybrid}
 what were later called
 hybrid inflation models.
 The models  were rather complicated, in part
because of a demand that the initial condition for observable inflation
is to be set by  an era of thermal equilibrium.

It was gradually recognized that prior thermal equilibrium is not  necessary.
A second  strand of model-building, characterized by little contact
with particle physics and focusing exclusively on non-hybrid models,
 began with the  proposal of chaotic inflation \cite{chaotic}.
Considerable attention was paid to
non-Einstein gravity theories, notably the proposal
of Extended Inflation \cite{extended}. In its original form that
 proposal is not
viable if the inflaton perturbation generates the curvature perturbation
\cite{ll92},
though it becomes viable if the curvature perturbation is generated afterward
\cite{mn}.

Following the formulation of a  simple  hybrid inflation
model \cite{andreihybrid},  attention went back to the connection with
particle physics and supersymmetry.
Almost all proposals for field theory beyond the Standard Model were considered
as arenas for inflation model-building, including especially GUTs and the
origin of low-energy supersymmetry breaking.

The most recent phase of model-building, beginning in about 2000,
is  based directly on  brane world scenarios. 
 We will consider the prediction of these kind of models 
 without describing their string-theoretic derivation. 

\section{Beyond the Standard Model}

We begin with some general ideas about the very early Universe, taking on
board  current thinking  about what may lie beyond the
Standard Model of particle physics.
Guided by the desire to generate primordial perturbations from the
vacuum fluctuation of scalar fields, one usually supposes
 that an effective
four-dimensional (4-d) field theory applies after the observable Universe
leaves the horizon,
though not necessarily with Einstein gravity.

To generate perturbations from the vacuum fluctuation we need $|aH|$
to increase with time, which is achieved by  inflation defined
as an era of  expansion with $\ddot a>0$ (repulsive gravity).\footnote
{As usual $a(t)$ is the scale factor of the Universe and $H\equiv \dot a/a$
is the Hubble parameter.}
 Perturbations  would also be generated from the vacuum
during an era of contraction with $\ddot a<0$
The original suggestion was called the
pre-Big-Bang \cite{gv}.
 A more recent version where the
bounce corresponds to the collision of branes was called the ekpyrotic
Universe \cite{ekpyr},
 which was further developed to produce a
cyclic Universe \cite{cyclic}.  In these scenarios, the prediction for the
perturbation depends crucially on what is  happens  at the bounce,
which is presently unclear.

Returning to the inflationary scenario, the 4-d field theory which is
supposed to be valid from observable inflation onwards cannot apply
back to an indefinitely early era.
The point at which it breaks down is a matter of intense debate at
present.  With Einstein gravity, 4-d field theory cannot be valid
if the energy density exceeds
 the Planck scale $\mpl\equiv (8\pi G)\mhalf=2.4\times 10^{18}\GeV$.
This is because quantum physics and general relativity
come into conflict at that scale, making it the
era when classical spacetime first emerges.
More generally, it   is supposed that any field theory will be just an
effective one, valid when relevant energy scales
are  below some `ultra-violet
cutoff' $\Lambda$. Above the cutoff, the field theory will be replaced either
by a more complete field theory, or  by a completely
different theory which is generally assumed to be   string theory.

 The measured values of the
gauge couplings suggest the existence of a GUT theory, implying
that field theory holds at least up to $10^{16}\GeV$.  This 
has not prevented the community from considering the possibility
that field theory fails at a much lower energy. The idea is that
4-d spacetime would emerge as an approximation to
the 10-d spacetime within which string theory is supposed to hold.
String theory is formulated in terms of fundamental strings (F strings),
but nowadays an important role is supposed to be played by what are 
called  D-$p$ branes (or just D branes) with various space dimensions 
$0<p\leq 9$.  The
electromagnetic, weak and strong forces that we experience might be
 confined to a particular D-3  brane,
 while gravity is able to penetrate to the region outside known as the
 bulk. An important role may be played by D strings, which are D branes
with just one of our space dimensions.


\section{The initial condition  for observable inflation}

\label{s8init}

The models of inflation that we are going to  consider apply to at least
the last
50 $e$-folds or so, starting with the exit from the horizon of the
observable Universe. One may call this the era of observable
inflation, because it is directly constrained by observation through
the perturbations which it generates. Assuming Einstein gravity,
observable inflation has to take place with energy density
$\rho\lsim (10\mtwo\mpl)^4$ or primordial gravitational would have
been detected. 

The era before observable inflation
is not directly accessible to observation,  but one may still
 ask about that era. In particular one may ask how the inflaton
field arrives at the starting point for observable inflation.
Though not compulsory, it normally is imagined that inflation begins
promptly with the emergence of 4-d  spacetime.  This is indeed desirable for
two reasons. One is to prevent the observable
Universe from collapsing
if the density parameter
$\Omega$ is initially bigger than 1 (without being fine-tuned to a
value extremely close to 1).
The other, which applies also to the case $\Omega\,{<}\,1$, is that
inflation protects an initially homogeneous region from invasion by
its inhomogeneous surroundings. This is because the event horizon
which represents the farthest distance that an inhomogeneity can
travel, is finite during inflation.
If the onset of  inflation is significantly delayed,
 one would need either a huge initially
homogeneous patch or \cite{andreiper}  a periodic universe.
In contrast, if inflation begins promptly with the emergence of 4-d
spacetime,  the initially homogeneous region is safe provided only
that it is bigger than the event horizon.  For
almost-exponential inflation the event horizon is of order the Hubble
distance.

A simple hypothesis about the emergence of 4-d spacetime
was made   in  \cite{chaotic}.
Working in the context of Einstein gravity, the energy
density of the Universe at the Planck scale is supposed to be
dominated by scalar fields, with the potential in some regions of
order $\mpl^4$ and flat enough for inflation to occur there.
This setup was termed chaotic inflation, and as an example the
potentials $V(\phi)\propto \phi^2$ and $\phi^4$ were considered.
These are generally called chaotic inflation potentials, but the
proposal of \cite{chaotic} regarding the initial condition
 does not rely on a specific form for the potential.
It {\em is} necessary though that there are regions of field space where
the potential is at the Planck scale and capable of inflating. No
example of such a potential has been derived from string theory.

An alternative to the chaotic inflation
proposal is that inflation begins at the top of a hill
in the potential, whose height  is much less than $\mpl^4$. In
particular, the height could be $\lsim (10^{16}\GeV)^4$, allowing observable
inflation to take place near the hilltop. This proposal is viable
even if the process by which the field arrives at the hilltop
is very improbable (such as the process of quantum tunneling through a
potential barrier), because inflation starting sufficiently near the hilltop
gives what is called { eternal inflation}
\cite{eternal1,eternal2}.

During eternal inflation, the volume of the inflating region grows
indefinitely, and it can  plausibly be argued that this indefinitely
large volume outweighs any finite initial improbability.  
Taking into account the
quantum fluctuation, it can be shown \cite{ks} that
eternal inflation takes place near a hilltop  provided
that $|\eta|<6$ where $\eta\equiv V''/3H^2$.

Eternal inflation near a hilltop  has been called
topological eternal inflation \cite{topological}. More
generally, eternal inflation occurs whenever the potential over a
sufficient  range satisfies
\be
\( \frac{H^2}{2\pi\dot\phi\sub{class}} \)^2 =
 \frac1{12\pi^2} \frac{V^3}{\mpl^6 {V'}^2} > 1
\label{eternalcon}
.
\ee
Here $\dot\phi\sub{class} = -V'/3H$ is the slow-roll approximation,
excluding the stochastic \cite{stochastic}
quantum fluctuation $H/2\pi$ per Hubble time.
  When the left-hand side of \eq{eternalcon}
is bigger than 1 the fluctuation dominates so that it can
overcome the slow-roll behaviour for an indefinitely long time, during
which eternal inflation occurs. In the opposite regime, the
fluctuation is small and the left-hand side of \eq{eternalcon} becomes
the spectrum of the curvature perturbation. Eternal inflation occurs
with the chaotic inflation potential $V\propto \phi^p$, for
sufficiently-large field values \cite{chaoticetern}.

Eternal inflation provides a realization of the multiverse idea, according
to which all possible universes consistent with fundamental theory (nowadays,
string theory) will actually exist \cite{eternal2,chaoticetern}.
 This is because eternal inflation can be of indefinitely long duration,
allowing time for tunneling to all local minimal of the scalar field potential.

\section{Slow-roll inflation}

\subsection{Basic equations}

We will find it useful to classify the models
 according to the  variation $\Delta\phi$
of the inflaton field after the observable Universe leaves the horizon.
We will call a model  small-field if $\Delta\phi \ll    \mpl$,
medium-field if $\Delta\phi\sim \mpl$ and
and  large-field  if $\Delta\phi \gg \mpl$.
Hybrid inflation models 
are usually constructed to be of the small-field type, the idea being to make
close contact with particle physics which is hardly possible for
medium- and large-field models.

The inflaton  field equation is
\be
\ddot\phi + 3H\dot\phi + V'(\phi) = 0 .
\ee
Except near a maximum of the potential (or minimum in the case of hybrid
inflation) a significant amount of  inflation can hardly  occur unless
this equation is  well-approximated by
\be
3H \dot \phi \cong  -V'
\label{phidot}
,
\ee
with  the energy density $3\mpl^2H^2=V+\frac12\dot\phi^2$
slowly varying on the Hubble timescale:
\be
\dot H \ll H^2 \label{slowh}
.
\ee
\eqs{phidot}{slowh} together define the slow-roll approximation, and we
will use $\cong$ to denote equalities which become exact in that approximation.

Consistency of \eq{phidot}
with the exact equation requires
\be
3\mpl^2 H^2 \cong V
\label{vofh}
\,.
\ee
and the flatness
conditions
        \be
            \epsilon\ll 1 \qquad  |\eta|\ll 1 \label{flat}
\,,
        \ee
   where
\be
\epsilon \equiv  \frac12\mpl^2(V'/V)^2 \label{epsilondef} 
\qquad \eta \equiv   \mpl^2V''/V \label{etadef}
\ee
Requiring that successively higher derivatives of the
 two sides of \eq{phidot} are equal to good  accuracy gives
more flatness conditions involving more slow-roll parameters.
The first two are
\bea
|\xi^2|& \ll& 1,{\qquad} \xi^2 \equiv
\mpl^4\frac{V'(d^3V/d\phi^3)}{V^2}
\label{xidef}\label{srxi}, \\
|\sigma^3| &\ll & 1,{\qquad} \sigma^3\equiv
\mpl^6\frac{V'^2(d^4V/d\phi^4)}{V^3} \label{sigdef}
.
\eea
The general expression is
\be
|\beta_{(n)}^n|\ll 1,{\qquad} \beta^n_{(n)}\equiv
\mpl^{2n}\frac{V'^{n-1}(d^{n+1}V/d\phi^{n+1})}{V^n} \label{betadef}
,
\ee
but only    $\xi^2$ and $\sigma^3$ are ever invoked in practice.

It is obvious that these additional parameters can have
 either sign.
The motivation for writing them as powers
comes from some simple forms for $V$,
which make  $|\xi|$, $|\sigma|$ and $|\beta_{(n)}|$  at most of order
$\eta$. For more general potentials one can check  case-by-case how small are
$\xi^2$ and $\sigma^3$. Usually there is at least a  hierarchy
\be
 \eta \gg \xi^2 \gg \sigma^3 \cdots
\label{hierarchy} ,
\ee 
but slow-roll {\em per se} requires only that all of the slow-roll parameters
are $\ll 1$ and does not require any hierarchy.

A convenient time variable is $N(t)$, the number of $e$-folds of expansion
occurring after some initial 
time, given by $dN=-Hdt$. In the slow-roll approximation
\bea
H' &\cong& -\epsilon H \label{dh} \\
\epsilon'  &\cong& 2\epsilon (2\epsilon -\eta) \label{depsilon} \\
\eta' &\cong& 2\epsilon\eta - \xi^2,
\label{deta} \\
\xi'  &\cong& 4\epsilon\xi^2-\eta\xi^2 - \sigma^3
\label{dxi},
\eea
and so on, where a prime denotes $d/dN$. 
The first relation says that almost-exponential occurs. The
second relation says that
 $\epsilon$ varies slowly. Slow-roll does not guarantee that the 
 other  parameters are slowly varying, though this is guaranteed in
the usual case that the hierarchy \eqreff{hierarchy} holds.

The flatness conditions  are obtained by successive
differentiations of the  slow-roll approximation. Strictly speaking,
a  differentiation  might incur large errors so that $\eta$ or higher
 slow-roll parameters  fail to be small (compared with 1).
In practice though one expects  at least
the first few slow-roll parameters to be small.

\subsection{Number of $e$-folds}

To obtain the predictions, one needs the scale $k(\phi)$ leaving the
horizon when $\phi$ has a given value. 
The number of $e$-folds from then until the end of slow-roll
inflation at $\phi\sub{end}$ is
        \be
            N(k) \cong \mpl^{-2} \int^{\phi}_{\phi\sub{end}}
\(
\frac{V}{V'}
\)\diff \phi =\mpl\mone \left| \int^{\phi}_{\phi\sub{end}}
\frac{\diff\phi}
{\sqrt{2\epsilon(\phi)}} \right| .
            \label{nofk}
        \ee
For definiteness we will evaluate the predictions for the
biggest cosmological scale $k=a_0H_0$, where the subscript
0 denoted the present epoch, and denote $N(a_0H_0)$ simply by
$N$.
The prediction for any other scale can be obtained using
\be
N(k)   = N- \ln(k/H_0) \equiv N - \Delta N(k) \label{deltan}.
\ee
Taking the shortest cosmological scale to be the one enclosing mass
$M=10^4\msun$, those scales span a range  $\Delta N = 14$.

The value of $N$ depends on the evolution of the scale
factor after inflation. With the  maximum inflation scale 
 $V\quarter= 10^{16}\GeV$ and radiation domination from inflation
onwards,  $N=61$. Delaying reheating until $T\sim \MeV$,
with matter domination before that, reduces this by 14.
With the maximum inflation scale it is therefore reasonable to adopt as an
estimate
\be
N= 54 \pm 7
\label{Nest3}
\,,
\ee
 Reducing the inflation scale reduces $N$ by $\ln(V\quarter/
10^{16}\GeV)$, and  the lowest scale usually considered is $10^{10}\GeV$
or so, reducing the above central value to 40.

Based on this discussion it seems fair to say that the fractional
uncertainty in $N$ is likely to be at most of order $20\%$.
As we shall see, the corresponding
uncertainties in the predictions are of the same order in a wide range of
models.
On the other hand, a very low inflation scale  and/or
Thermal Inflation \cite{thermal}
could reduce $N$ by an indefinite amount. The only absolute constraint
is  $N>14$, required  so that perturbations are generated on all cosmological
scales.  Also, a long era of domination by the kinetic term of a scalar field
(kination), corresponding to $P=\rho$, could  increase the
estimate \cite{andrewn} by up to 14. Taking all of that on board the maximum
range would be $14<N<75$.

In non-hybrid models, $\epsilon$ usually increases with time and
inflation ends when one of the flatness conditions fails, after which $\phi$
goes to its vacuum expectation value (vev). {}From its definition,
 $\epsilon$ increasing  with time corresponds to
$\ln V$ being  concave-downward. In this case, the
 value of $\phi_*$ obtained from \eq{nofk} will typically be insensitive to
$\phi\sub{end}$, making the model more predictive.

 In some
 hybrid models, $\epsilon$ decreases with time
($\ln V$ concave-upward), and inflation ends only when the
waterfall field is destabilized. In other hybrid inflation models
though,
 $\epsilon$ increases with time ($\ln V$ concave-downward),
and {\em slow-roll}
inflation may end before the waterfall field is destabilized through the
failure of  one of the flatness conditions.
If that happens,  a few  more $e$-folds of inflation can take place
while the inflaton  oscillates about its vev (locked inflation \cite{locked}),
until the amplitude of the oscillation becomes low enough to destabilize the
waterfall field.

\subsection{Predictions}

The vacuum
fluctuation of the inflaton generates a
practically gaussian perturbation, with spectrum
$\calp_\phi(k) = (H_k/2\pi)^2$ where the subscript $k$ indicates horizon
exit $k=aH$. This perturbation  generates
 a  time-independent curvature perturbation
with  spectrum \cite{specpred}
\be
\calp_\zeta(k) = \frac1{24\pi^2\mpl^4} \frac {V_k}{ \epsilon_k}
\label{spec}
.
\ee

The error in this estimate will come from the error in $\calp_\phi$
and the error in the slow-roll approximation. Both are expected to 
give a small fractional error,  of order $\max\{epsilon,\eta\}$.
Differentiating with respect to $\ln k$ to get 
the spectral index may incur a fractional error $\gsim 1$ if $\eta$
is rapidly varying  \cite{ewanref}, but that is not the case 
in the usual models. Differentiating  \eq{spec} using 
\eqs{dh}{depsilon} give  the spectral tilt;
\be
n-1\equiv d\ln \calp_\zeta/d\ln k =2\eta_k-6\epsilon_k . \label{npred}
\ee
If in addition $d\eta/dN$ (equivalently, $\xi^2$)
 is slowly varying this may be differentiated
again to obtain the running,
\be
dn/d\ln k =  -16\epsilon\eta +24\epsilon^2 + 2\xi^2.
. \ee

Observable inflation can take place near a maximum or minimum of the potential
even with the  flatness condition $|\eta|\ll 1$ mildly violated
to become $|\eta|\sim 1$ (so-called fast-roll inflation \cite{fastroll},
though note
that $\dot\phi$ is still small making $H$ almost constant).\footnote
{Very close to a maximum is the regime of eternal inflation,  which
presumably precedes fast- or slow-roll inflation.}
This quite natural possibility would give tilt $|n-1|\simeq 1$, which is also
quite  compatible with the original arguments of
Harrison \cite{harrison}  and Zeldovich \cite{zeldovich}
 for $n\sim 1$ and all known
environmental arguments. The very small tilt now observed is not
required by any general consideration, and a large tilt $n-1\sim -0.3$
had previously  been  considered as a serious possibility to make a
critical-density CDM model more viable \cite{ll92}.

During inflation, the vacuum fluctuation generates a primordial tensor
perturbation, setting the initial amplitude for gravitational waves
which oscillate after horizon entry. The spectrum $\calp\sub T$
of this perturbation is conveniently specified by the
 tensor fraction $r\equiv \calp\sub T/\calp_\zeta$. In
the slow-roll approximation
 \cite{ll92},\footnote
{The definition of $r$ in this reference was slightly different.}
\be
r = 16\epsilon = -8 n\sub T
\label{rbound}
,
\ee
where  $n\sub T\equiv d\ln \calp\sub T/d\ln k$.
The second relation has become known as the
consistency condition, and its violation would show that the
curvature perturbation is not generated by a  single-field slow-roll
inflation.

Using the observed value for the spectrum of the curvature perturbation,
the tensor fraction is given by
\be
r = \( \frac{V\quarter}{3.3\times 10^{16}\GeV} \)^4
\label{rofv} .
\ee
The tensor fraction  can also be related to  $\Delta\phi$.
Suppose that slow-roll persists to almost the end of inflation and that
$\ln V$ is concave-downward throughout. Then
 $|V/V'|$ is continuously increasing,
and  \eq{nofk}  gives $2\epsilon < N^{-2}( \Delta\phi/\mpl)^2 $.
This can be written \cite{mygrav,bl}
\be
16\epsilon = r<  0.003 \( \frac {50}N \)^2 \( \frac{\Delta\phi}\mpl \)^2
\label{rbound06} .
\ee

Now suppose instead that slow-roll persists to the end of inflation,
without any requirement on the shape of the potential. As a consequence
of slow-roll,  $\epsilon$
varies little during one Hubble time and there are only 50 or so Hubble
times. It follows that one may expect $\epsilon$ to
 be at least roughly constant, in which case
the right hand side of \eq{rbound06} provides
at least a
 rough estimate of the actual value of $\Delta\phi$.

Finally, let us adopt the most conservative possible position and consider
just the change $\Delta\phi_4$ during the four $e$-folds after the observable
Universe leaves the horizon, that being the era when an observable
 tensor perturbation may 
actually be generated. Then it is certainly safe to assume that  $\epsilon$
has negligible variation, leading to the quite firm estimate
\be
r\simeq \frac12 \( \frac{\Delta\phi_4}{\mpl} \)^2
\label{rbound1}
.
\ee

        \begin{figure}
 \centering\includegraphics[angle=270,width=1.0
\columnwidth,totalheight=2.5in]{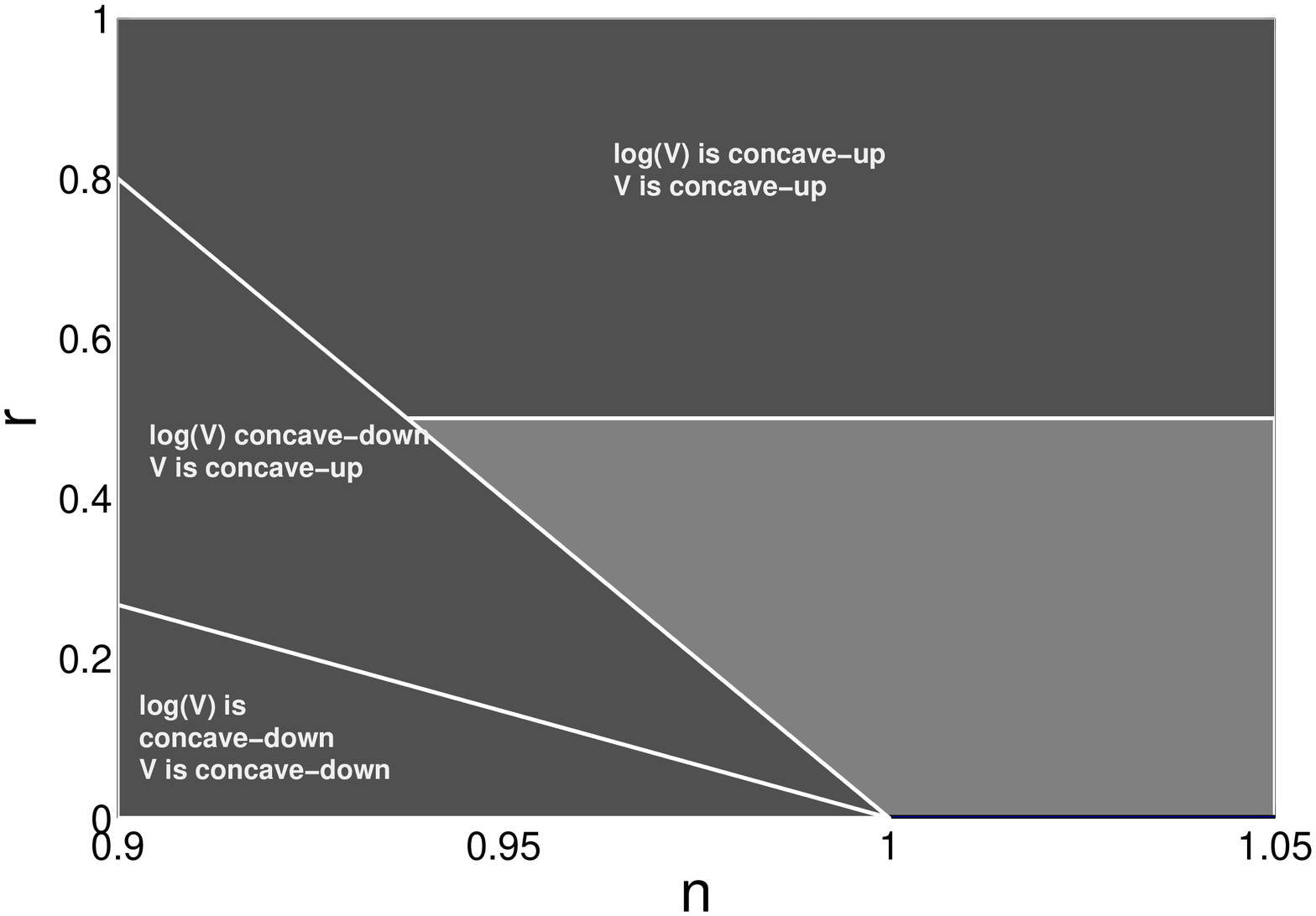}
        \caption{}
    \label{rlinear}
 \end{figure}

        \begin{figure}
        \centering
\includegraphics[angle=270,width=0.9\columnwidth,totalheight=2.5in]{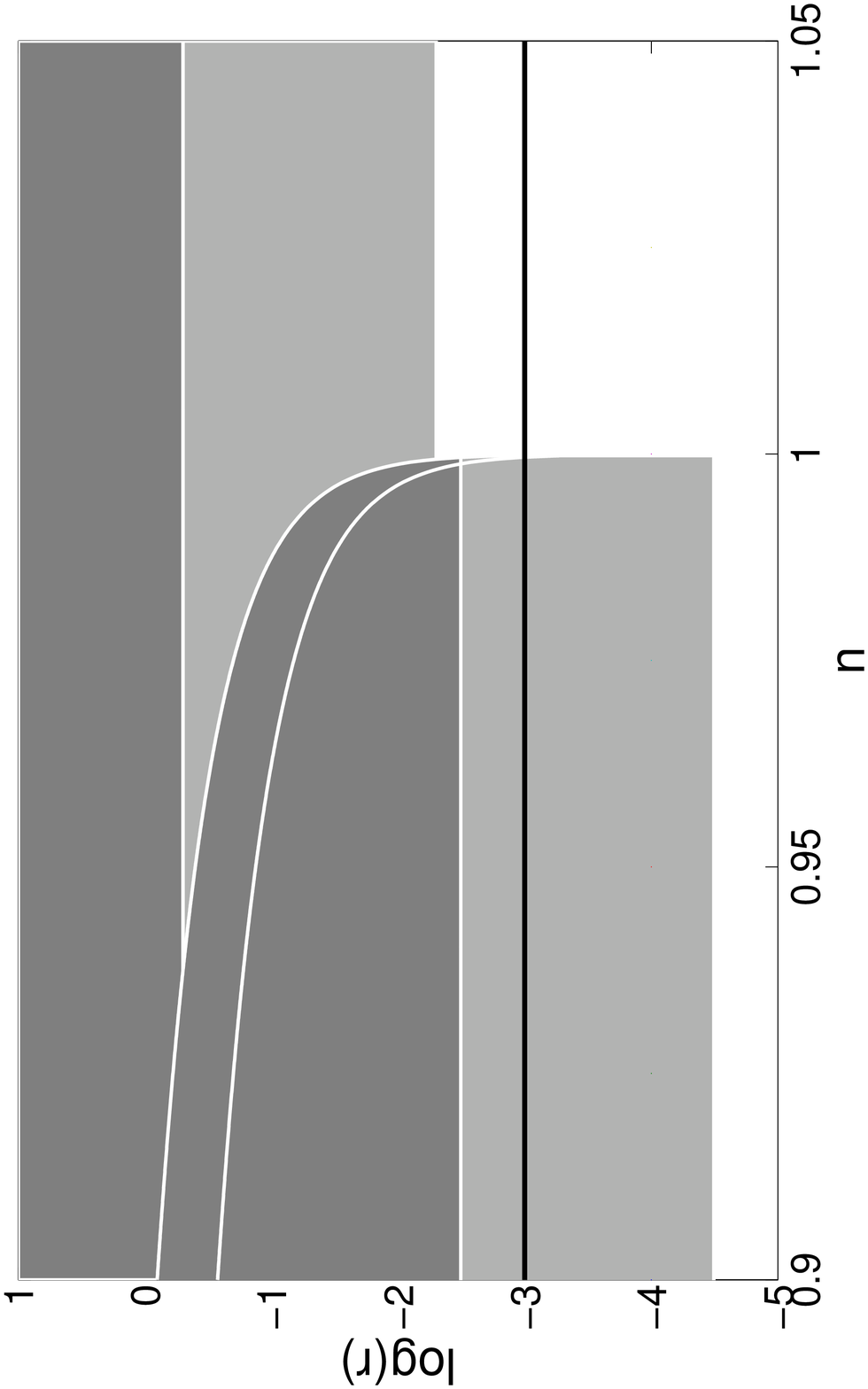}
            \caption{}
        \label{rlog}
        \end{figure}

In Figure  \ref{rlinear}, the $r$-$n$ plane is 
 divided  into three regions, according to whether $V$
and $\ln V$ are concave-upward or concave-downward while cosmological scales
leave the horizon. Figure \ref{rlog} repeats the plot in the $\ln r$-$n$
plane.

 If the concave-upward -downward behaviour
persists till the end of slow-roll inflation, the right-hand region is
inhabited exclusively by hybrid
 inflation models,  since otherwise inflation would never end.
With that assumption, \eqs{rbound06}{rbound1}  imply that  the
lightly-shaded region of the Figures is excluded 
 if $\Delta\phi>0.1\mpl$, and that
the  heavily-shaded region  region is excluded if $\Delta\phi>\mpl$.
(In the right-hand region, corresponding to concave-upward $\ln V$,
we used \eq{rbound1} with $\Delta\phi_4=\Delta\phi$; the actual bound will be
tighter since in reality $\Delta\phi_4<\Delta\phi$.)

\subsection{Observational constraints}

According to observation  \cite{wmap}
value of the spectrum  $\calp_\zeta$ has the almost scale-invariant
value $(5\times 10^{-5})^2$, with negligible error.
This gives the constraint
\be
V^{1/4}/\epsilon^{1/4} = 0.027 \mpl = 6.6 \times 10^{16}\GeV,
\label{vbound}
\ee
which we will call the cmb constraint.

 Setting $r=0$ and taking $n$ to be scale-independent, observation
gives  \cite{wmap}
$n\simeq  0.948^{+ 0.015}_{-0.018}$.  Allowing $r$ and a scale-independent
$dn/d\ln k$ gives a higher $n$ and $n'\simeq -0.10\pm 0.05$, consistent with
no running at $2\sigma$ level.
The allowed region in the $r$-$n$ plane  is shown in
Figure \ref{rvsn}. (This is  a corrected version of the Figure in \cite{wmap},
kindly supplied by the authors). The  bound $r=0$  is seen to apply
for $r\ll 0.1$.\footnote
{The 1-$\sigma$ limit with $r$ set equal to zero is tighter
than the limit read off from setting $r=0$ in the $r$-$n$ plot, because
the joint probability distribution is non-gaussian}.
Within
a few years  there will be either a detection of $r$ or a bound $r<10\mtwo$.
If  $r$ is below $10^{-3}$ it  will probably be undetectable by any means.
This value is marked  in Figure \ref{rlog}.

        \begin{figure}
        \centering
        \includegraphics[angle=270,width=1.0\columnwidth]
{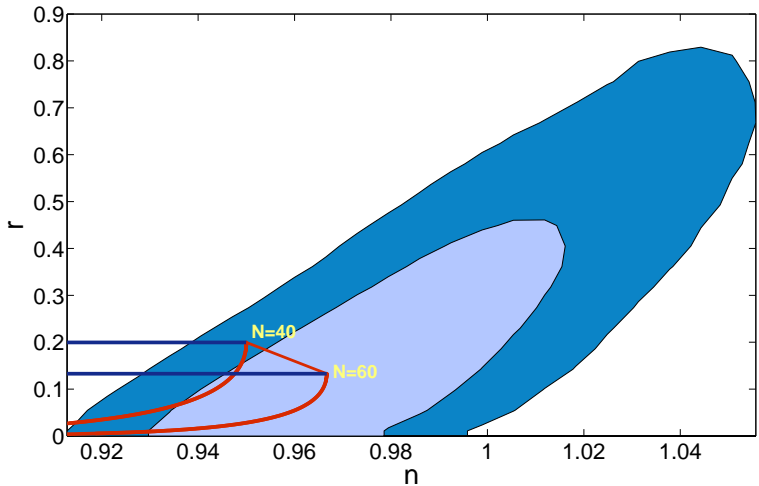}
        \caption{The closed areas show the regions
 allowed by observation at $66\%$ and $95\%$ confidence levels.
The curved lines are the Natural Inflation predictions for
$N=20$
   and $N=75$, and the horizontal lines are the corresponding multi-field
        Chaotic Inflation predictions. The junction of each pair of lines
corresponds         to single-field Chaotic Inflation.}
                                    \label{rvsn}
                                \end{figure}

{}From all this, we see that small- and 
 medium-field generally give   $r\lsim 10\mtwo$. This means
that the predicted tensor fraction is unlikely to be observed.
It also means that the prediction for the spectral tilt can 
be taken as  simply $n-1=2\eta$; to reproduce the observed negative
tilt  the potential of a
small- or medium-field model should be 
 concave-downwards
while cosmological scales leave the horizon.

\subsection{Beyond the standard paradigm}

\label{sgenpred}

Throughout we have adopted the standard paradigm, whereby
 the curvature perturbation $\zeta$ is generated
by the inflaton perturbation in a single-field slow-roll inflation
model. In general there will exist other light fields, each possessing
a perturbation with the nearly flat spectrum $(H/2\pi)^2$,
any one of which might be responsible for the curvature perturbation.

The predictions in this more general scenario are best calculated
through the $\delta N$ formalism \cite{starob85,ss1,lms,lr,st}.
 As our main
 focus is on the standard paradigm we just give some basic results
without derivation. It
is convenient to use at horizon exit a field basis $\{\phi,\sigma_i\}$,
where $\phi$ points along the inflaton trajectory and the $\sigma_i$
($i=2\cdots M$) are orthogonal. The perturbation $\delta\phi$
then generates the same time-independent curvature perturbation as in the
single-field case, whose spectrum we denote by
$\calp_{\zeta_\phi}$.
 The orthogonal  perturbations give no contribution
to the curvature at horizon exit, but one or more of them
may generate an additional  contribution later  which may be
dominant by the time that the curvature perturbation settles down to the
final time-independent
value (obtaining as cosmological scales start to approach the horizon)
whose spectrum we denote simply by $\calp_\zeta$.
The additional  contribution may be generated during inflation in which case we
are dealing with a multi-field inflation model, or later through for
example the curvaton mechanism \cite{curvaton}. In the latter case,
the model of inflation is irrelevant; all that matters is that the Hubble
parameter is slowly varying. Liberated from the constraint to generate the
curvature perturbation, model-building becomes much easier \cite{dl}.

The cmb normalization \eqreff{vbound}
now becomes an upper bound, implying a lower inflation scale for
a given value of $\epsilon$.
The spectral index in general depends on the
evolution after horizon exit \cite{ss1,treview}, but in the most natural
case that the contribution
of single orthogonal field
$\sigma\equiv \sigma_i$
dominates it is given by the potential at horizon exit as
\be
n(k)-1 = 2 \eta_{\sigma\sigma} - 2\epsilon
,
\ee
where $\eta_{\sigma\sigma}= \pa^2 V/\pa\sigma^2$.
(The case that  two contributions are comparable may
 arise  by accident, or in special models where $\phi$ and an
orthogonal field are related such as the one involving axion physics
which is described in \cite{george}.)

Since
the  tensor perturbation
depends only on $H$ the tensor fraction $r$ is reduced;
\be
r = 16\epsilon  \frac{\calp_{\zeta_\phi}}{\calp_\zeta}
 <  16\epsilon
.
\ee
It is negligible if an orthogonal contribution dominates.

We did not mention  non-gaussianity. 
According  to the standard paradigm, the non-gaussianity is \cite{maldacena}
 about 100 times smaller than the level that
can be detected from the cmb anisotropy (and/or galaxy surveys) though
it has recently
been claimed \cite{cooray} that a measurement from the 21-cm anisotropy might
be possible. In contrast, non-standard paradigms may easily
generate  non-gaussianity at an observable level; in particular the
 curvaton and inhomogeneous reheating scenarios are 
expected to generate non-gaussianity at a level that is at least marginally
observable  through the cmb.  If non-gaussianity is
observed we will be dealing with functions (of  rotationally-invariant
scalars formed from the wave-vectors that define  the bispectrum, trispectrum
etc.) as opposed to numbers, 
which will provide powerful information about the origin of the
curvature perturbation.

All of this assumes slow-roll inflation.
That possibility   is compatible with the simultaneous detection
of a   tensor perturbation {\em and}  non-gaussianity only
if some  orthogonal field can generate the non-gaussianity
without being dominant (a highly constrained scenario \cite{bl2}).
The main alternative to slow-roll inflaton seems to be
  inflation with
non-quadratic kinetic terms, called
$k$-inflation \cite{kinflation},  of which  special forms are  the
brane world DBI inflation scenario
\cite{dbi} and ghost inflation \cite{ghost}.

\section{Modular inflation}

We begin our survey of inflation models
with the most plausible  medium-field model,
which goes by the name of modular inflation. This is a  non-hybrid model
in which the inflaton is a modulus. It was
  suggested a long time
ago \cite{bg}
 and its possible realization in
the context of brane worlds is under investigation at present.

Moduli may play other roles too in the
early Universe, and we describe their  properties before getting to the
inflation model.
For the present purpose a modulus may be defined as a field with a  potential
of the form
\be V= V_0  f(\phi/\mpl)
\label{modpot05}
, \ee
This is supposed to hold  in the range $0<\phi\lsim \mpl$,
with the function $f(x)$ and its low derivatives of order 1 at a generic
point. At the vev, where $f$ and $f'$ vanish, the mass-squared  $m^2\equiv
V''$ is typically of order $V_0/\mpl^2$. If the potential has a maximum,
it will typically be located at a distance of order $\mpl$ from the vev
with the  tachyonic  mass-squared $V''$ typically of order $-m^2$.

Fields with this property are expected (though not inevitable)
in a field theory derived from string theory. Usually the field theory
is taken to be supersymmetric though moduli are expected anyway.
Moduli are usually  supposed to have interactions of only gravitational
strength, corresponding to a lifetime $\Gamma\sim m^3/\mpl^2$.
Alternatively though, a modulus may have interactions of ordinary strength,
in particular gauge interactions. The fixed point of the symmetry group
is then called a point of enhanced symmetry. Such a point
 might correspond to either the vev or to a maximum
of the potential. It may even be  possible for both of these to be
points of enhanced symmetry, involving  different symmetry groups.

Moduli may affect cosmology in several ways. Usually they are considered in
the context of supersymmetry, and the simplest expectation for the mass
is then
$m \sim \TeV$, corresponding to what we may call light moduli.
A light modulus is typically displaced strongly from its
vev during inflation, by an amount which puts
its subsequent oscillation and gravitational-strength decay into conflict with
nucleosynthesis. To avoid this
`moduli problem' one may suppose that all moduli are heavy, or that there is
Thermal Inflation \cite{thermal}.

Now we turn to modular inflation. It is usually supposed to take place near a
maximum or saddle-point of the potential, with just one modulus $\phi$
varying significantly.
As many moduli  typically exist, that   may  not be easy to arrange.
Supposing that it happens let us set $\phi=0$ at the maximum and consider
the  power series for the potential. The  generic expectation would be  for the
quadratic term alone to provide at least a crude approximation to the
potential in the slow-roll regime, corresponding to
\be
V(\phi) = V_0 \( 1 + \frac12 \eta_0 \frac{\phi^2}{\mpl^2} \)  \label{vhill}
\ee
But this requires  (from \eq{modpot05}) roughly $\eta_0\sim -1$
which gives spectral tilt $n-1\sim -1$ in contradiction with observation.
To provide a modular inflation model one suppresses
the quadratic term,   either by means of a symmetry \cite{rs} or more usually
by fine tuning (see for instance \cite{modular2}).

If the suppressed 
quadratic term is still required to dominate while cosmological scales
leave the horizon, one obtains the scale-independent prediction
$n=1+2\eta_0$  which can agree with observation by choice of  $\eta_0$.
This prediction is scale-independent which might in the future allow it
to be distinguished from other predictions for $n$. Of course, one
has to invoke additional terms to end inflation, presumably at a value
$\phi\sub{end}\sim \mpl$.  The tensor fraction is
\be
r=2\( \frac{\phi\sub{end}}{\mpl} \)^2 (1-n)^2 e^{-N(1-n)}
\sim 10^{-3.5} \( \frac{\phi\sub{end}}{\mpl} \)^2
.
\ee
Taking  $\phi\sub{end}\sim \mpl$ gives the result shown in
Figure \ref{newrvsn}. The tensor fraction is 
unobservable,  but corresponds to a high normalization scale
$V\quarter\sim 10^{15}\GeV$, meaning that we 
 are not dealing with a light modulus.

It is more reasonable to suppose that the suppressed quadratic term is
negligible. 
Then, as a rough approximation it may be reasonable
to write
\be
V\simeq V_0 \[  1- \( \frac{\phi}\mu \) ^p \]
,
\label{higher}
\ee
\noindent with $p\gsim  3$ (not necessarily and integer)
 and $\mu\sim\mpl$.

If  this approximation holds  for some reasonable length of
time after cosmological scales leave the horizon it  gives
\be
\phi*^{p-2}=\[ p(p-2)\mu^{-p} N \mpl^2 \]^{-1} ,
\label{phiend}
\ee
(independently of $\phi\sub{end}$) and
\be
n-1 = -\frac 2 N \left(\frac{p-1}{p-2}\right)
.
\label{ncubhigh}
\ee
For the range  $3<p<\infty$ with $N=50$ we get $0.92<n<0.96$.
The cmb  normalization corresponds to a tensor fraction
\be
r \simeq \frac{0.001}{(p-2)^4} \( \frac \mu\mpl \)^\frac p{2p-4}
 \( \frac{50}N \)^\frac{2(p-1)}{p-2}
\label{rcubhigh}
.
\ee
This is shown in Figure \ref{newrvsn} 
with $\mu=\mpl$. Again, the tensor fraction
 is too small to detect but still   corresponds to a
 high energy scale $V\quarter \sim 10^{15}\GeV$.
These estimates agree to rough order of magnitude with results obtained
numerically using potentials derived from string theory (see for instance
\cite{modular2}).

\section{Small-field models}

A range of small-field models has been proposed. Before describing them
we make some general remarks, followed by a  very basic treatment
of supersymmetry  which is invoked in  most small-field  models.

The motivation for small-field  models comes from
 ideas about what is likely to lie beyond the Standard Model of
particle physics.
 Choosing the origin as the fixed
point of the relevant symmetries, the tree-level
potential will have a  power series
expansion,
\be
V(\phi) =V_0 \pm \frac12 m^2\phi^2 + M\phi^3
+\frac14\lambda\phi^4+ \sum_{d=5}^\infty \lambda_d \mpl^4
\( \frac\phi\mpl \)^d
\label{power}
\ee
 The lower-order terms of \eq{power}, which do not involve
$\mpl$, are   renormalizable  terms (corresponding to a renormalizable
quantum field theory). The
higher-order terms, which disappear in the limit $\mpl\to \infty$, are
 non-renormalizable terms.
We are taking $m^2$ positive and as indicated the quadratic term
might have  either sign. The other renormalizable terms will usually be 
positive, but the non-renormalizable terms might have either sign.

According to
a widely-held view, non-renormalizable terms of arbitrarily high order are
 expected,  with magnitudes big
enough to place this expansion out of control at $\phi\gsim \mpl$.
The typical expectation is $|\lambda_d|\sim  1$ if $\mpl$ is the ultra-violet
cutoff and  $|\lambda_d| \sim(\mpl/\Lambda)^d$ 
(the latter corresponding to the replacement
$\mpl\to \Lambda$) if the cutoff $\Lambda$ is smaller.
This view is part of a more general one, according to which the lagrangian
of a field theory ought to contain all terms that are allowed by the
symmetries, with coefficients typically of order 1 in units of the ultra-violet
cutoff (see for instance \cite{weinberg1}).

If the field theory is replaced by a more complete field theory
 above the cutoff,
 the $\lambda_d$ can be calculated and  will be of
the advertised order of magnitude if $\phi$ has unsuppressed interactions.
But if instead it is replaced by  string theory above the cutoff,
then  estimates of $\lambda_d$ should come from  string theory.
Such estimates are  at present not available, except for
moduli.\footnote
{In the case of moduli \eq{modpot05}
implies a strong suppression of the couplings. However, the inflaton
in a small-field model is not usually supposed to be a modulus because
the origin in small-field models is usually taken to be the fixed point of
the symmetries of some unsuppressed interactions, which  would make the origin
a point of enhanced symmetry for the modulus.}
In general then, one is free to accept or not the usual view about
non-renormalizable terms.\footnote
{This is less true if supergravity is invoked
 because  the non-renormalizable terms are then present and out of control
for generic choices of the functions defining the theory. But one can
still make special choices to avoid the problem.}

Following \cite{treview},
let us see what sort of conditions the terms in \eq{power} must satisfy,
to achieve inflation in the small-field regime $\phi\ll\mpl$.
We discount the possibility of extremely accurate cancellations between
different terms.  This means that the  constant term has to
dominate, and that we require the addition of any  one
other term to respect the flatness condition
 $|\eta|\ll 1$, the other flatness conditions then being automatic.\footnote
{In a supersymmetric theory  one instead consider $A$-term inflation
 \cite{aterm1,aterm2}.
Dropping the constant term $V_0$, one can choose a flat
direction (say in the space of the MSSM scalars) in which
the leading non-renormalizable term in the superpotential generates an
 $A$-term. Then a fine-tuned match between three terms in the potential
 can give $V'=V''=0$ 
for a particular field value. Inflation can then take place
near that value and naturally reproduce the cmb normalization.
 By a suitable choice of the fine-tuning it can also reproduce the 
observed spectral index, though it can also give 
 any value in the slow-roll
range $0\lsim n\lsim 2$  \cite{aterm2}.}

We shall not
consider the cubic term, which usually is forbidden by a symmetry. For
the other terms $|\eta|\ll 1$ is equivalent to
\bea
m^2 &\ll& \frac{V_0}{\mpl^2 } \simeq 3 H_*^2  \label{etaprob} \\
\lambda &\ll& \frac{V_0}{\mpl^4}\frac{\mpl^2}{\phi^2}\label{lambdacon} \\
\lambda_d &\ll& \frac{V_0}{\mpl^4}
\( \frac{\mpl^2}{\phi^2} \)^{\frac{d-2}2} \label{lambdadcon}
.
\eea
One might think that the second and third conditions can always be
satisfied  by making $\phi$ small enough, but this is not correct
because there is  a lower limit on the variation of  $\phi$.
Indeed, during just the ten or so $e$-folds while
cosmological scales leave the horizon
  \eqs{nofk}{vbound}  require $\phi$ to change by at least
$ 10^4  V\half/\mpl$ and $\phi$ cannot be smaller than that
on all such scales.  We conclude that
\bea
\lambda &\lsim & 10^{-8}\label{lambdacon2} \\
\lambda_d &\lsim & 10^{-8} \( \frac{ 10^{16}\GeV}{V_0^{1/4}} \)^{2(d-4)}
\label{lambdadcon2}
.
\eea

The first  condition requires $\lambda$ to be very small, and the second
condition
requires at least the first few $\lambda_d$ to be very small unless the
inflation scale is well below $10^{16}\GeV$.
Supersymmetry can ensure these
conditions, either by itself or combined with an internal symmetry.
Alternatively one can invoke just an internal symmetry corresponding
to $\phi\to \phi+$\,const, making $\phi$ a PNGB, though as we remark later
that is not so easy to arrange as one might think.

Finally, we recall  that for a generic field in an effective field
theory, $\mpl$ in \eq{power} might be replaced by an ultra-violet
cutoff $\Lambda\sub{UV}< \mpl$, arising either because heavy fields
have been integrated out, or because large extra dimensions come into
play.  One hopes  that such a thing does not happen for the inflaton
field, because it would make it more difficult to satisfy the flatness
conditions \cite{myextra}.
 Fortunately, the presence of large extra dimensions does
not in itself prevent $\mpl$ from being the effective cutoff for at
least some of the fields.

\section{Supersymmetry: general features}

Field theory beyond the Standard Model is usually required to possess
supersymmetry.
Supersymmetry \cite{weinberg3} is an extension of Lorentz invariance.
Its outstanding prediction is that each  fermion should have  bosonic
superpartners, and vice versa, with identical mass and couplings in the
limit of unbroken supersymmetry.  Supersymmetry has to be broken 
in our Universe.

Supersymmetry is usually
taken  to be a local symmetry,
and is then  called { supergravity} because it
automatically incorporates gravity.\footnote
{Some brane world scenarios explicitly break local supersymmetry
which means there is actually explicitly broken global supersymmetry.}
In that case the breaking is spontaneous.
 In many situations, global
supersymmetry is used with the expectation that it will provide a good
approximation to supergravity.  In that case the breaking can be 
spontaneous and/or explicit. 

We shall deal with the simplest
version of supersymmetry,
known as $N=1$ supersymmetry, which alone seems able to provide
a viable extension of the Standard Model.
 Here, each spin-half field is paired with either a
complex spin-zero field (making a { chiral supermultiplet}), or
else with a gauge boson field (making a { gauge supermultiplet}).
With supergravity, 
 the graviton (spin two) comes with a gravitino (spin $3/2$).
With spontaneously broken global supersymmetry there is instead a spin $1/2$
goldstino.

One motivation for supersymmetry concerns the mass of the Higgs
particle, given by the vev of $\pa^2 V/\pa \phi^2$ where $\phi$
is the Higgs field. The function $V$ that we have up till now being calling
simply the potential is only an effective one, and not the `bare'
potential  entering into
the lagrangian which defines the field theory.
Interactions of the scalar fields with themselves and
each other change the bare potential into an effective potential.
We will be concerned with perturbative quantum effects 
 represented by Feynman diagrams.
If we including just tree-level (no-loop) diagrams, the effective potential
is still given by the power series \eqreff{power} with different
(renormalized) values for the  coefficients in the
series.  Loop corrections give further renormalization
of the coefficients,  which is our immediate concern. (They also
give the potential  logarithmic terms that have to be added to the power
series, which we come to later.)

The point now is that  the loop `correction' in
 a generic field theory will be large, driving the physical mass up to a value
of order the ultra-violet cutoff.
As the latter is usually
supposed to be many orders of magnitude above the physical Higgs mass,
one must in the absence of supersymmetry fine-tune the bare mass so that it
almost exactly cancels the loop  correction.
To protect the Higgs mass from this fine tuning, 
one needs to keep the loop  correction under
control by means of a symmetry which would make it zero
in the unbroken limit. The best symmetry for doing that job in the case of
the  Higgs field is supersymmetry.\footnote
{If a symmetry other than supersymmetry were to be used,
 the Higgs field $\phi$ would become a PNGB corresponding to a shift
symmetry $\phi\to \phi+\,$constant. It is difficult for a shift symmetry
to protect the Higgs mass, because the symmetry will be broken by the
strong couplings that the  Higgs is known to possess.
 This problem can be overcome
by what is called the Little Higgs mechanism but the resulting schemes
are complicated especially if the ultra-violet cutoff is supposed to be
many orders of magnitude bigger than the observed mass.}

In a  supersymmetric extension of the
Standard Model, each particle species
must come with a superpartner.  It turns out
that at least two Higgs fields are then needed. Keeping just two, one
arrives at the Minimal Supersymmetric Standard Model (MSSM), which is
a globally supersymmetric theory with canonically-normalized fields.
The partners of the quarks and leptons are called squarks and
sleptons, those of the Higgs fields are called higgsinos, and those of
the gauge fields are called gauginos.

Unbroken supersymmetry would require that each Standard Model
 particle has the same mass as
its partner.  This is not observed, which means that the global supersymmetry
possessed by the MSSM must be broken in the present vacuum.  To agree
with observation it turns out that the breaking has to be explicit as
opposed to spontaneous.   
 To ensure that supersymmetry
continues to do its job of stabilizing the potential against loop
corrections, the breaking must be of a special kind call soft
breaking.  Soft supersymmetry breaking has  to give slepton and squark
masses very roughly of order $100\GeV$.  They cannot be much smaller
or they would have been observed, and they cannot be much more bigger
if supersymmetry is to do its job of stabilizing the Higgs mass.

Softly broken  supersymmetry explains with
high accuracy the observed ratio
of the three gauge couplings (determining the strengths of the
 strong, weak and electromagnetic interactions) on the hypothesis that there
is a GUT. This feature is actually preserved if one allows the squarks and
sleptons
to be extremely heavy (hence not observable), a proposal known as Split
Supersymmetry.

The LHC will soon determine the nature of the fundamental
interactions immediately beyond the Standard Model, and may or may not
find evidence for supersymmetry. In the latter case we will know
that supersymmetry is too badly broken to be relevant for the Standard
Model. It  might still be relevant in the early Universe and in
particular during inflation, but there is no doubt that increased
emphasis will then be placed on non-supersymmetric inflation models.
A good candidate for non-supersymmetric inflation would
be modular inflation.  Alternatively, one might
make  the inflaton a PNGB, or just accept extreme fine tuning.

\section{Supersymmetry: form of the  potential}

In a supergravity theory, the  potential is a function of the complex
scalar fields, of the form
\be
V(\phi_i) = V_+(\phi_i) - 3\mpl^2 m_{3/2}^2(\phi_i)
\dlabel{vsugra07}
. \ee
The first term is positive, and spontaneously breaks supersymmetry.

In the vacuum, $m_{3/2}(\phi_i)$ becomes the gravitino mass which we denote
simply be $m_{3/2}$. Let us denote the vev of the first term by $M\sub S^4$.
The near-cancellation of the two terms in the vacuum is unexplained (the 
cosmological constant problem). The explicitly broken global supersymmetry 
seen in the MSSM sector is supposed to be obtained from the full
potential as an approximation. To achieve this the spontaneous breaking must
take place in some `hidden sector' with some `messenger' sector
communicating (mediating) between the hidden sector and the MSSM sector.
The value of $M\sub S$ required to give 
 squark and slepton masses of order   $100\GeV$ depends on the strength
of the mediation. Let us characterize it by $M\sub{mess}$, with
$100\GeV = M\sub S^2/M\sub{mess}$. Gravitational-strength mediation
(`gravity mediation') corresponds to $M\sub{mess}\sim\mpl$ 
and the biggest reasonable
range is $10^4\GeV \lsim M\sub{mess} \lsim 10^{12}$.\footnote
{The upper limit corresponds to  anomaly mediation, which is
gravity mediation suppressed by a loop factor. The lower limit  
is an interpretation of $M\sub S\gg 100\GeV$, required so that the hidden
sector is indeed hidden.}
The corresponding gravitino mass is between $1\eV $ and $10^6\GeV$.

Coming to inflation, supersymmetry stabilizes  the potential against loop
corrections just as in the MSSM Higgs case.  Also, the 
small   $\lambda$  required in the tree-level potential can be obtained
quite naturally. One  generally assumes that the first term
of \eq{vsugra07} dominates since there is no reason to expect a fine 
cancellation. Assuming that  supersymmetry in the early Universe is
broken at least as strongly as in the vacuum, this requires 
$V\gsim M\sub S^4$. Partly for that reason,  very low-scale inflation 
is difficult to achieve.

Now we come to what has been called the $\eta$ problem.
The supergravity potential can be written as the sum of two terms,
called the $F$ term and the $D$ term. In most inflation models $V$
comes from the $F$ term. Then, each scalar field typically has 
mass-squared  at least of order  $m^2\gsim V/\mpl^2=3H^2$.
 For the inflaton this is  
 in mild conflict with the slow-roll requirement $|\eta|\ll 1$
\cite{os83,coughdine,cllsw,ewansugra}.

Even if we allow the curvature perturbation to be generated 
after inflation, say in the curvaton model,
we still need $m^2\ll V/\mpl^2$ for the curvaton  field.
In that case
 there may be a problem even after inflation, because
a generic supergravity theory  still gives each scalar
field an effective   mass at
least of order $H$ \cite{drt} except during radiation
domination \cite{lm},  which will tend to drive
each field to its unperturbed value and kill the curvature perturbation.

Returning to the standard scenario for generating the curvature perturbation,
 we  typically need $|\eta|\sim 0.01$ to
generate the observed spectral tilt. This represents  an order  one
percent fine-tuning which is not too severe. What is perhaps
more serious is that
the  $\eta$ problem calls into question the validity of any model which
 is formulated within the context of global supersymmetry. It is easy to ensure
$|\eta|\ll 1$ in such a theory, but having done that the supergravity
correction  may still be big and completely alter the model.
In a typical global supersymmetry model though, the same is true of other
types of correction  as well.

\section{One-loop correction}

\label{s:7.8.3j}

Loop corrections add a logarithmic term to the
effective potential.  In the direction of any field $\phi$, the one-loop
correction is
\be
\Delta V(\phi)\,{=}\,
\sum_i \frac{\pm\kern.7pt{\cal N}_i}{64\pi^2} M^4_i(\phi)
        \ln\left[\frac{M_i^2(\phi)}{Q^2} \right].
\label{loop1}
\ee
This is called the  Coleman-Weinberg potential. The sum
goes over all particle species, with the plus/minus sign for
bosons/fermions, and ${\cal N}_i$ the number of spin states. The
quantity $M^2_i(\phi)$ is the effective mass-squared of the species,
in the presence of the constant $\phi$ field.  For a scalar,
$M_i^2=\pa^2 V/\pa
\phi_i^2$, which is valid for $\phi$ itself as well as other scalars.

The quantity $Q$ is called the { renormalization scale}. If the
loop correction were calculated to all orders, the potential would be
independent of $Q$. In a given situation,
 $Q$ should be set equal to a typical energy
scale so as to minimize the size of the loop correction and its accompanying
error.  Focusing on the inflaton
potential, we should set $Q$ equal to a typical value of $\phi$ (one
within the
range which corresponds to horizon exit for cosmological scales). That
having being done, the {\em magnitude} of $\Delta V$ will typically be
negligible, but its derivatives may easily be significant.

If supersymmetry were unbroken, each spin-1/2 field would have a scalar- or
gauge field partner with the same mass and couplings, causing the loop
correction to vanish.  In reality supersymmetry is broken. To see how things
work out, let us consider the loop correction from a chiral
supermultiplet, consisting of a spin-1/2 particle with a scalar
partner.  The partner is a complex field $\psi=(\psi_1 +
i\psi_2)/\sqrt 2$, whose real components $\psi_i$ have true masses
$m_i$.  If there is an interaction $\frac12
\lambda' \phi^2 |\psi^2|$, this gives $M_i^2 = m_i^2 +\frac 12\lambda'
\phi^2$ ($i=1,2$).  (We use the prime to distinguish this coupling
from the self-coupling $\lambda$ in the tree-level potential
\eqreff{power} of the inflaton.)
The spin-1/2 field  typically has true mass $m\sub
f\,{=}\,0$, and  its interaction with $\phi$ generates an
effective mass-squared $M\sub f^2(\phi) =
\frac12\lambda' \phi^2$.
 (This result is not affected by either
spontaneous or soft supersymmetry breaking.)  When $\phi$ is much bigger
than $m_i$, the loop correction is therefore
\be
\Delta V \simeq \frac1{32\pi^2 } \left[ \sum_{i=1,2} \left(m_i^2
+ \frac12 \lambda'\phi^2\right)^2 - 2\left(\frac12\lambda'
\phi^2\right)^2 \right]\: \ln \frac{\phi}{Q}.
\label{ajf_8.8}
\ee

The coefficient of $\phi^4$ vanishes by virtue of the supersymmetry.
For the  other terms, we will consider two cases. Suppose first 
that global supersymmetry is spontaneously broken 
during inflation. Then it turns out that  typically
$m_1^2 =- m_2^2$, causing the  coefficient of $\phi^2$ in Eq. (\ref{ajf_8.8})
to vanish. This leaves 
\be
\Delta V \simeq \frac{m_1^4}{32\pi^2} \ln \frac{\phi}{Q}.
\label{spont}
\label{vdval}
\ee
In this case the derivatives of $\Delta V$ are independent of $Q$, making its
choice irrelevant as the magnitude of $\Delta V$ is negligible.

Now suppose instead that 
global supersymmetry is explicitly (softly) broken during inflation, the
coefficient of $\phi^2$ in \eq{ajf_8.8} does not vanish, but instead
typically dominates the constant term. Adding the loop correction to
the mass term of the tree-level potential gives
\be
\Delta V= \frac12 \left[m^2 + \frac{\lambda'}{32\pi^2}
\( m_1^2 + m_2^2\) \ln \frac{\phi}{Q} \right] \phi^2,
\label{vrunning}
.
\ee
This expression is valid over a limited range of $\phi$, if $Q$ set
equal to a value of $\phi$ within that range. If a large range of
$\phi$ is under consideration, it should be replaced by an expression
of the form
\be
\Delta V=\frac12 m^2(\phi) \, \phi^2
.
\label{vrun98}
\ee
\noindent The  ``running mass'' $m^2(\phi)$ is calculated
from what are called renormalization group equations (RGEs).

The above discussion involved the loop correction due to a chiral
supermultiplet.  Couplings involving chiral super multiplets, such as
$\lambda'$, are called { Yukawa couplings} and they can be very
small.  We could instead have discussed the loop correction due to a
gauge supermultiplet, consisting of a spin-1/2 field whose partner is
a gauge field.  The couplings involving gauge super multiplets are
called { gauge couplings} and denoted usually by $g$.  They are not
expected to be very small.  The loop correction from a gauge
supermultiplet is essentially of the above form, with $\lambda'$
replaced by $g$.

Finally, if
 there is no supersymmetry, the loop correction typically
destabilizes the tree-level potential, and in particular it gives to
the mass of each scalar field a contribution which is typically of
order the ultra-violet cutoff. To obtain an acceptable potential, and
in particular acceptable masses, one has to invoke a fine-tuned
cancellation between the loop correction and the tree-level potential.
 Considering just the contribution from
the spin-1/2 part of
\eq{ajf_8.8}, and adding it to the self-coupling of $\phi$,  one has
\be
\Delta V = \frac14 \[ \lambda -\( \frac{\lambda'}{4\pi} \)^2  \ln \frac\phi Q
\] \phi^4
.
\ee
As with the mass, the RGE's give  a more accurate  result, corresponding to
$\Delta V = \frac14 \lambda(\phi) \phi^4$
with a running coupling $\lambda(\phi)$.

\section{Small-field models:  moving away from the origin}

\label{saway}

 \begin{figure*}
   \begin{minipage}[t]{0.5\linewidth}
 \centering\includegraphics[angle=270, width=2.5in]{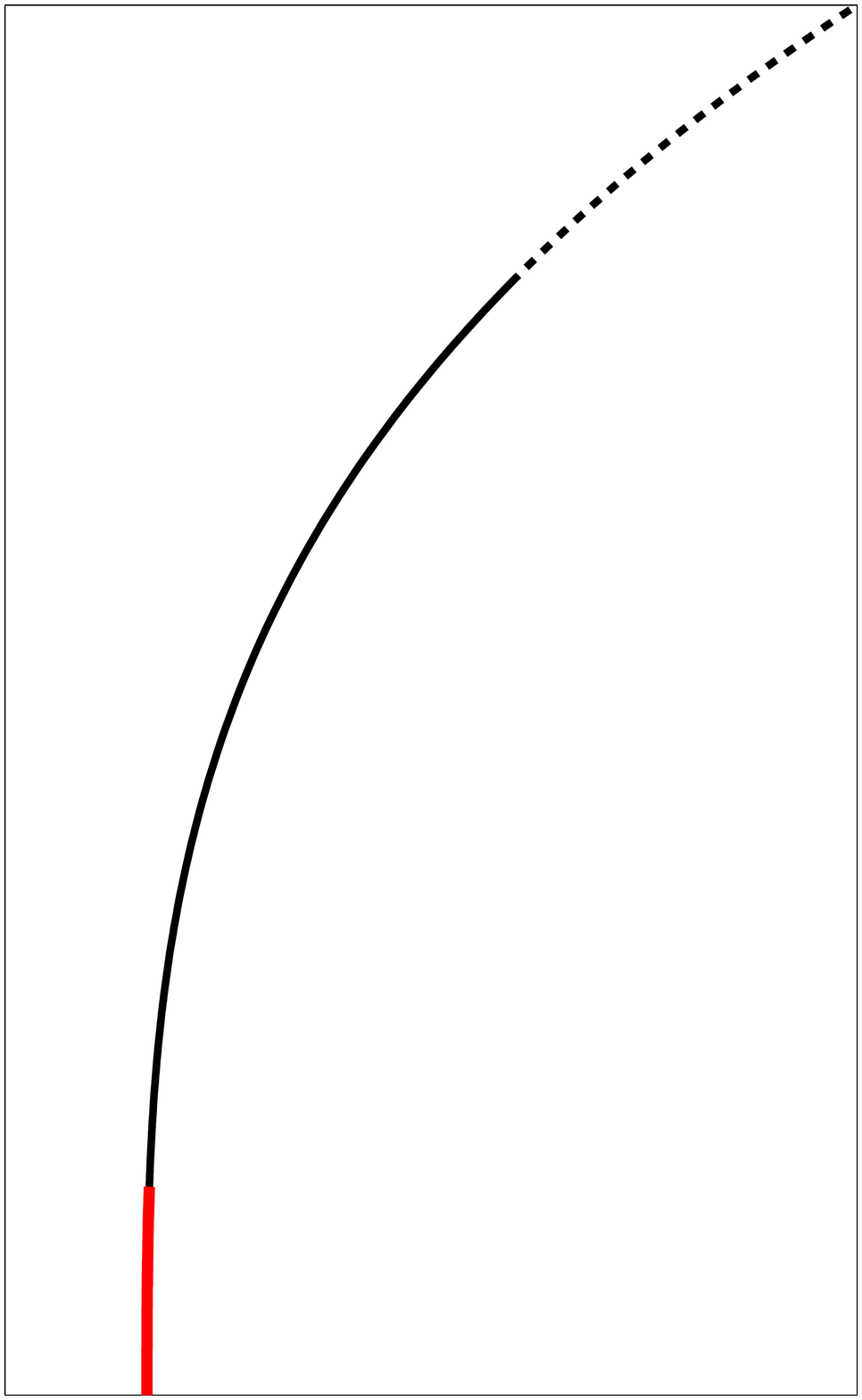}
  \caption{Modular, new, inverted hybrid, mutated hybrid.}
    \label{smallv1}
\end{minipage}%
   \begin{minipage}[t]{0.5\linewidth}
\centering\includegraphics[angle=270, width=2.5in]{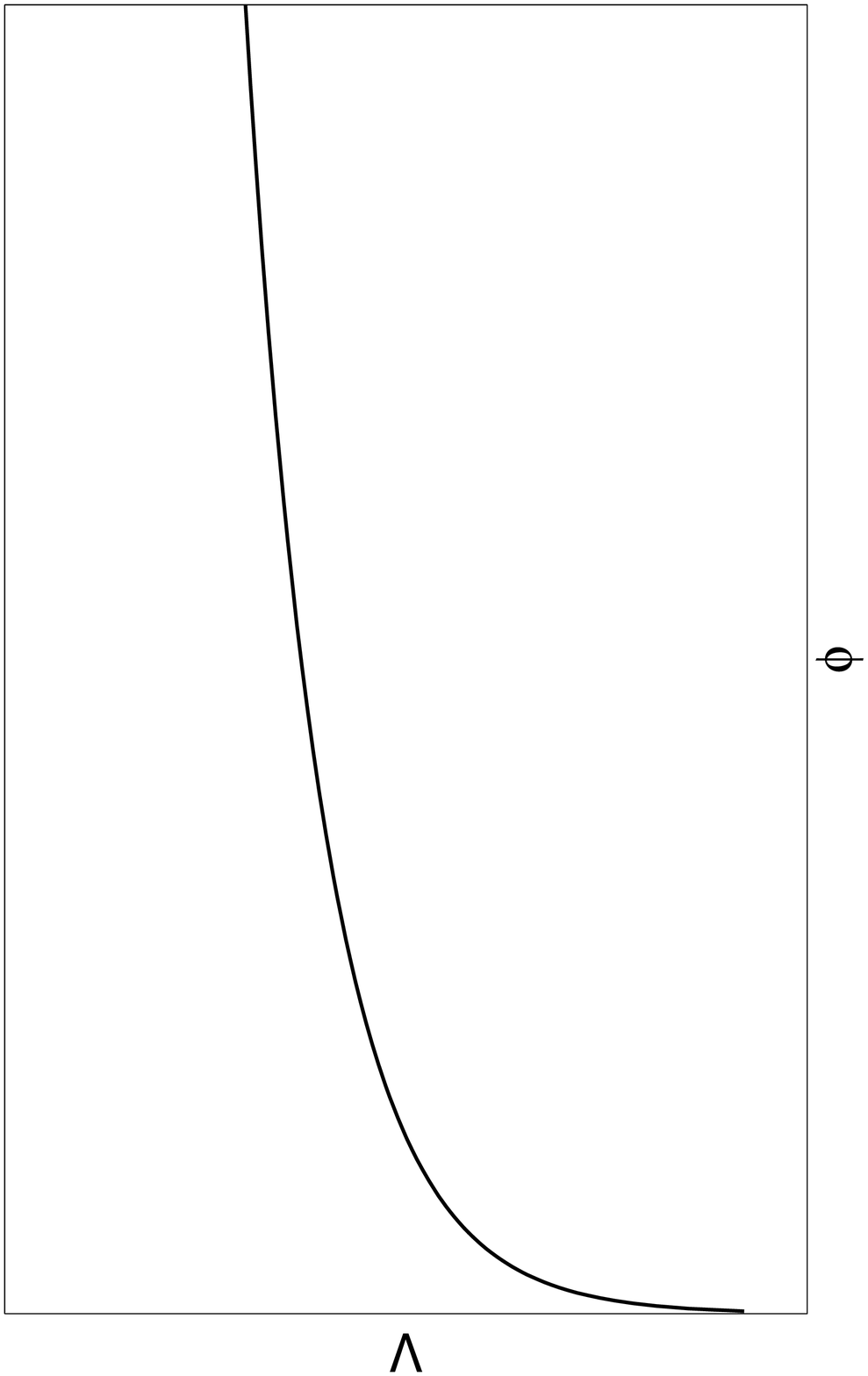}
\caption{$F$- and $D$-term inflation, colliding brane, mutated hybrid.}

                                    \label{smallv2}
                            \end{minipage}\\[20pt]
                            \begin{minipage}[t]{0.5\textwidth}
 \centering\includegraphics[angle=270,width=2.5in]{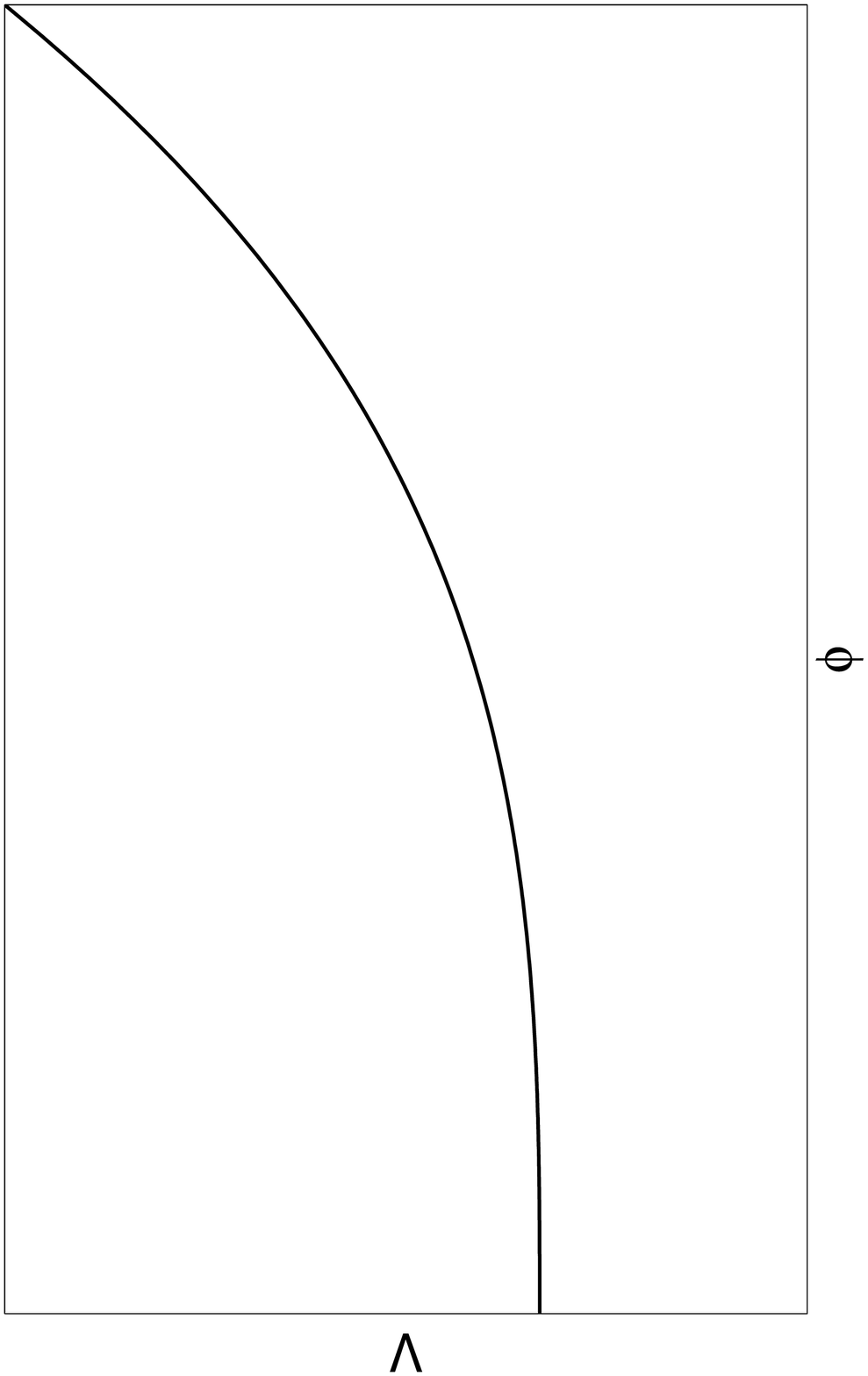}
                                \caption{Tree-level hybrid.}

                                    \label{smallv3}
                            \end{minipage}%
                            \begin{minipage}[t]{0.5\textwidth}
              \centering\includegraphics[angle=270,width=2.5in]{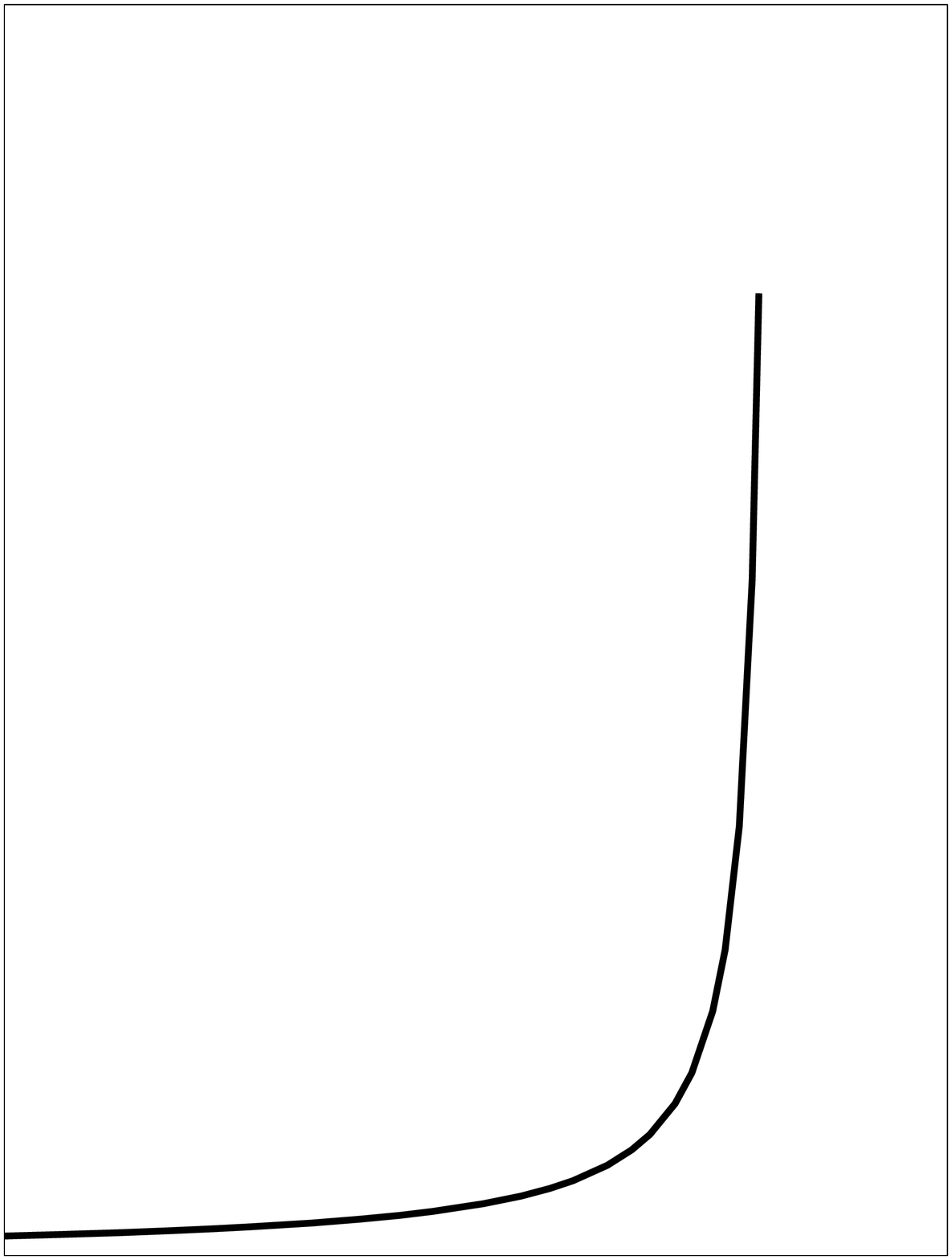}
           \caption{Dynamical supersymmetry breaking.}
\label{dbranes}
                            \end{minipage}\\[20pt]
                            \begin{center}
                            \begin{minipage}[t]{0.5\textwidth}
        \centering\includegraphics[angle=270, width=2.5in]{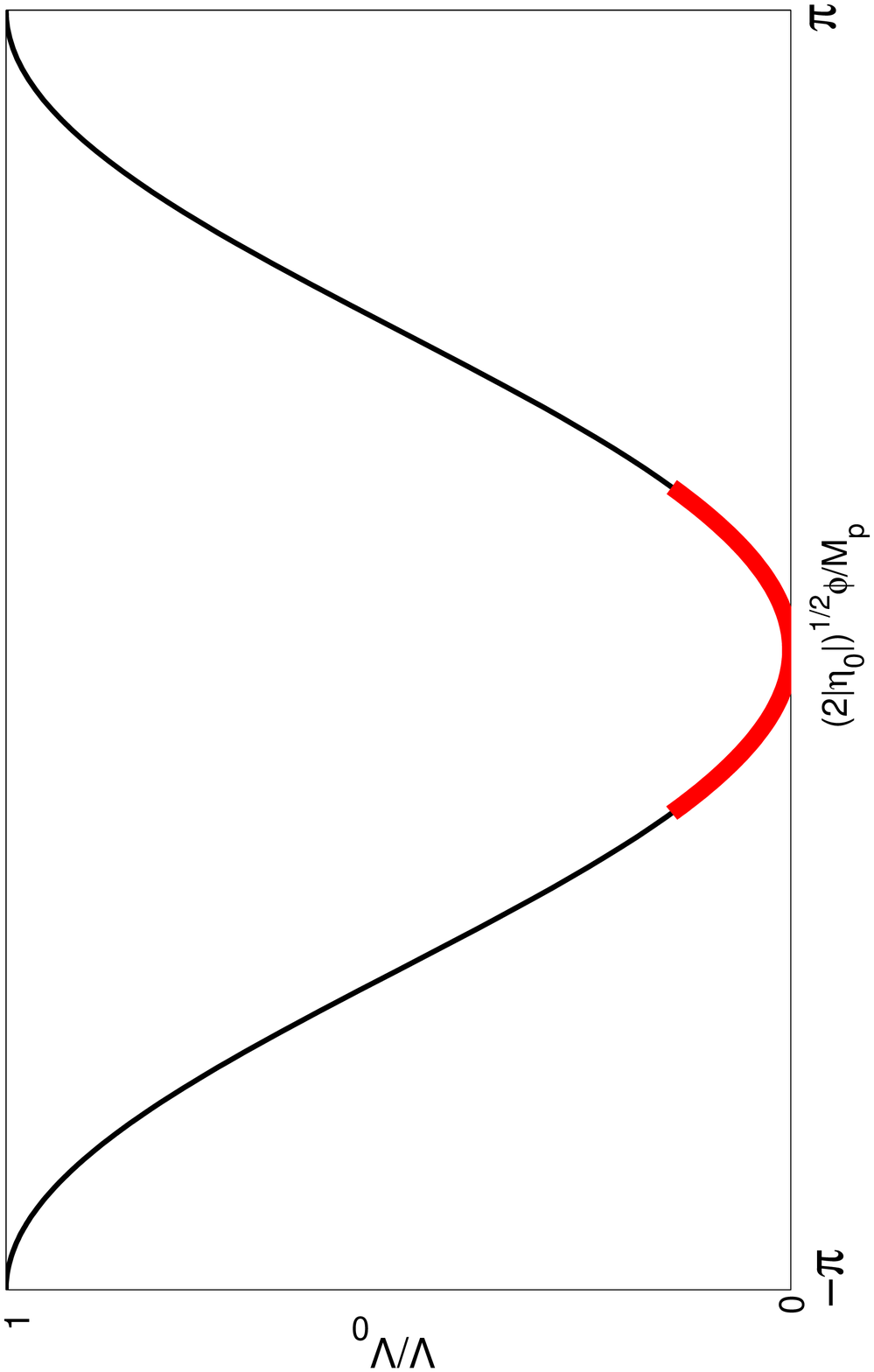}
                                    \caption{Natural/chaotic inflation}
                                    \label{sinusoidal}
   \label{smallv4}
                        \end{minipage}
                        \end{center}
                        \end{figure*}

In this section we consider small-field  potentials
with the shape shown in  Figure \ref{smallv1}.
We begin with  non-hybrid models, taking the origin as the fixed point
of the symmetries. Then the minimum of the potential corresponds to a
nonzero vev, and the potential vanishes there.
Such models
are usually called New Inflation models, since that was the name
given to the first viable slow-roll model which happened to be of that kind.

The situation for New Inflation is similar to the one we discussed for
modular inflation. Keeping the
quadratic term alone cannot be a good approximation throughout inflation.
Assuming that the quadratic term is already
negligible when  cosmological scales leave the horizon, the approximation
 \eq{higher} seems reasonable, with $p\gsim 3$ and now $\mu\ll\mpl$.
With this approximation the spectral tilt is  given by \eq{ncubhigh}.
The  tensor fraction is given by \eq{rcubhigh} with $\mu\ll\mpl$
making it absolutely negligible, and allowing an inflation scale
far below $10^{15}\GeV$.

The original New Inflation model corresponded to $p=4$;
\be
V \simeq  V_0 - \frac14\lambda \phi^4 + \cdots
\label{voflambda}
.
\ee
To be precise, the inflaton was supposed to be the GUT Higgs, taken to
be practically massless, whose Mexican-Hat potential was generated by
a running coupling coming from the non-supersymmetric Coleman--Weinberg
potential.
The cmb normalization now  requires
 $\lambda=3\times 10^{-13}(50/N)^3$. This ruled out the
model in its original form,  because  $\lambda$ was the GUT gauge coupling
with known magnitude of order $10\mone$. A viable version of the 
 model was obtained \cite{qaisar} by declaring that the inflaton is a
gauge singlet, making $\lambda$ a Yukawa coupling whose value can be 
chosen at will.

Instead of invoking the approximation \eqreff{higher},
 we might suppose that the quadratic
term dominates while cosmological scales leave the horizon but a
higher term dominates soon afterward.
The simplest potential of this kind is
\be
V= V_0 - \frac12m^2\phi^2 -\frac14 \lambda\phi^4 + \cdots
\label{simplest} .
\ee
 A supersymmetric realization of this case making close contact with particle
physics is given in \cite{dr} (see also \cite{treview}), which is
very fine-tuned if the inflaton is required to generate the curvature
perturbation.
There is also a
non-supersymmetric realization invoking a Little Higgs mechanism
\cite{pseudonat,kw},
 making
 $\phi$ a PNGB with a periodic  potential. 
The prediction for this model is the same as for \eq{vhill}, with the 
difference that $\phi\sub{end}$ will be far below $\mpl$ making the
inflation scale far below $10^{16}\GeV$. 

Turning to hybrid inflation, the simplest possibility is inverted hybrid
inflation \cite{inverted}
where the origin remains  the fixed point of symmetries, and
one simply reverses the sign
of $m^2$, $m_\psi^2$
 and $\lambda'$  in the usual  hybrid inflation potential (\eq{vord}
below).
The
negative sign of $\lambda'$ is difficult to arrange especially in a
supersymmetric model, and severe fine-tuning is also required
\cite{kingsan}.

Instead one can make $\phi$ a PNGB so that it has a periodic potential
\cite{cs,pseudonat,kw}.
The shift symmetry is broken both by the potential $V(\phi)$
and by the  coupling
of $\phi$  to the waterfall field. The inflationary trajectory does
not pass through the fixed point of the symmetries, and
taking the origin to be a maximum of the
potential is just an arbitrary choice.
Instead of making $\phi$ a true PNGB, one can arrange that at least
it is effectively one during inflation, in the sense that the potential
then becomes flat in some well-defined limit \cite{cllsw,ewansugra,glm}.
For both types of model it seems possible for the magnitude of the spectrum
and the spectral tilt to be in agreement with observation by suitable
choice of parameters.  The inflation scale
can be many orders of magnitude  below $10^{15}\GeV$.

 \section{Moving toward the origin; power-law potential}

In this section we consider potentials of the form illustrated in Figure
\ref{smallv2}, of either the small-field or medium-field type.
We begin with potentials that can be approximated by
\eq{higher} with $p<0$.
  Such potentials give the prediction
\eqreff{ncubhigh}
for the spectral index and  \eqreff{rcubhigh} for the tensor fraction.

With $p=-4$, \eq{higher} has been derived in a  brane world
scenario, where $\mu\sim\mpl$
is allowed corresponding to a medium-field model \cite{branereview}.  
This is a  hybrid inflation model, with the usual potential schematically
of the form
\be
V(\phi,\chi)  =  V(\phi) + \frac12m^2\phi^2 -\frac12 m_\chi^2\chi^2 +
\frac12\lambda '\chi^2\phi^2 +  \frac14\lambda  \chi^4  \label{vord}
.
\ee
At $\phi>\phi\sub c\equiv m_\chi/\sqrt{\lambda'}$ the waterfall field
is driven to zero, leaving $V(\phi)$ given by \eq{higher}. The unusually
form of $V(\phi)$ here arises because the  inflaton
field $\phi$ corresponds to the distance between  branes attracted 
towards each other. Inflation in this model 
ends when the branes coalesce.  

Colliding brane inflation has the usual $\eta$ problem, in that the potential
is expected to have a term $\frac12 m^2\phi^2$ with $m^2 \sim H^2$. But the
 brane world scenario can  motivate a non-canonical normalization of
a specific form,  leading to what is called DBI inflation which can
take place even with $m^2\sim H^2$. We shall not present the results for
that case.

At the end of this brane world inflation, F and  D strings are typically
produced. At present it is not clear how that affects the viability of the
model, because the evolution of the string network has not been reliably
calculated.

The potential \eqreff{higher} with various values of $p$ 
 had been derived earlier
 in the context of ordinary field theory,  with $\mu\ll\mpl$ corresponding
to a small-field model. The mechanism, referred to as mutated
\cite{mutated}
 or smooth \cite{smooth} hybrid   inflation, is the following.
The waterfall field is not fixed during inflation, but instead adjusts to
continually minimize the potential. The effective potential is then
$V(\phi,\xi(\phi))$, and for simplicity the $\phi$-dependence at fixed $\chi$
is taken to be negligible.
In this way  \cite{inverted} one can obtain any
$p<0$ (not necessarily integral)
as well as $p>1$. 
Taking negative $p$, the upper bound on $r$
(evaluated by setting $\Delta\phi <\mpl$) is shown in Figure
\ref{newrvsn}. 

This is a good place to mention another potential of the kind shown in
Figure \ref{smallv2};
\be
V\simeq V_0 \left[1- \exp \left(-q\frac{\phi}{\mpl}\right) \right],
\label{another}
\ee
with $q$ of order 1.  It occurs if inflation takes place in field
space where the kinetic function has a pole, irrespective of the form of the
potential \cite{ewansugra}, with model-dependent values of $q$ such as
$q=1$ or $\sqrt2$. It can also be obtained by transforming
 $R^2$ gravity or scalar-tensor gravity to the Einstein frame,
giving $q=\sqrt{2/3}$. Notice  that  these modified-gravity theories
should not be used in conjunction with the standard supergravity
potential,   because that potential is evaluated
in the Einstein frame.

The potential is supposed to apply in the regime where $V_0$
dominates, which is $\phi\,{\gsim}\,\mpl$.  Inflation ends at
$\phi_{\rm end}\sim \mpl$, and when cosmological scales leave the
horizon, we have $\phi \simeq \ln(q^2 N) \mpl/q$ and
\be
n \simeq 1+2\eta = 1 - \frac{2}{N}.
\ee
The predicted  cmb normalization (for $q=1$ and $N=50$)
is shown in Figure \ref{newrvsn} as a cross.

\section{$F$ and $D$ term inflation}

Now we suppose that the potential is dominated by  the
loop correction, in a   model invoking spontaneously-broken
global supersymmetry.
We focus initially on the case that the supergravity correction is
negligible, asking later whether that is reasonable in specific
models.   In the regime $\phi\gg\phi\sub c$ the potential  is then given by
\eq{spont}, while in the limit $\phi\to\phi\sub c$ it vanishes
(because $M_i(\phi)$ in \eq{loop1} vanishes). The mass-squared in
\eq{spont} is proportional to some coupling $g$ which controls the
strength of the spontaneous supersymmetry breaking.  The potential during
inflation is therefore of the form
\be
V(\phi) = V_0 \( 1+
\frac{g^2}{8\pi^2} f(\phi) \ln\frac \phi Q \)
\label{vsugra3}
,
\ee
where $f=1$ for $\phi\gg \phi\sub c$ and $f\to 0$ as $\phi\to\phi\sub c$.
The potential has the form shown in Figure \ref{smallv2}.

For $\phi\gg\phi\sub c$,
\be
\eta = -
\frac{ g^2}{8\pi^2} \frac{\mpl^2}{\phi^2} = - \epsilon \frac{\mpl}{\phi}
.
\ee

Consider first the regime
\be
g^2\gg 8\pi^2\phi\sub c^2/\mpl^2
\label{kappacon}
.
\ee
Slow-roll inflation ends at $\phi\sub{end} = 2g\mpl^2/4\pi\gg\phi\sub
c$, because $\eta=1$ there.  After {\em slow-roll} inflation ends,
$\phi$ oscillates about $\vev\phi=0$. A few $e$-folds (of order
$\ln (\phi\sub{end}/\phi\sub c)$) of `locked' inflation
then occur,  until the amplitude falls below $\phi\sub c$.

The integral \eq{nofk} is dominated by the limit $\phi$
 giving
\be
\phi \simeq \sqrt\frac{ N }{4\pi^2} \, g \mpl
.
\label{vv}.
\ee
To be in the desired regime $\phi\ll\mpl$ we need $g\ll 1$ which might
be in conflict with \eq{kappacon}. Proceeding anyway one finds
$n=1-1/N\simeq 0.98$, and the cmb normalization
$r= 0.0011 (50/N) g^2$. This prediction (with $N=50$)
is shown as a star in 
Figure \ref{newrvsn}.

All this is with $g$ in the regime \eq{kappacon}. If we decrease $g$
smoothly to reach the opposite regime
$g^2\ll   8\pi^2\phi\sub c^2/\mpl^2$,
$\phi(N)$ approaches $\phi\sub c$, the cmb normalization
decreases and $n$ approaches $1$ \cite{kq}.

Two versions of this model exist in the literature, referred to generally
as $F$-term  \cite{cllsw,ewansugra,dss}  and $D$-term 
\cite{ewandtof,dterm} inflation.\footnote
{The supergravity potential can be written as the sum of  an $F$ term and
a $D$ term. With the $D$ term one is driven more or less inevitably
to this type of model, but many other possibilities exist with the $F$
term.} In both cases, the starting point is a simple global 
supersymmetry theory with canonical kinetic terms, 
giving the hybrid inflation potential
\eqreff{vord} with $V(\phi)$ perfectly flat.

In the $F$-term case,  $g$ is a Yukawa coupling,
which can be chosen to be small yielding a small-field model.
The cmb normalization fixes the vev of the 
waterfall field, as $\Lambda\simeq
6\times 10^{15}\GeV$. Identifying the waterfall
field(s) as a subset of the GUT Higgs fields motivates
this value. Turning that around, the GUT model
predicts roughly the observed magnitude for the spectrum of the curvature
perturbation. 

As we are dealing with an $F$ term, the $\eta$ problem exists;
 we expect $V\simeq \pm m^2\phi^2$
with $m^2\sim H^2$. To have a viable model $m^2$ needs to be tuned down 
by a factor of order $0.01$ but there is no reason why it should be
negligible. 
The case of  positive $m^2$ has been investigated in  \cite{panag}
and negative $m^2$ in \cite{bl}. The latter case gives an attractive 
model because it corresponds to
 hilltop inflation as in Figure (\ref{logsketch}). After eternal inflation
near the hilltop,  the field can roll in the negative $\phi$
direction.
After re-defining the origin and reversing the sign of $\phi$ we recover
the small-field model considered in Section \ref{saway}. 
Taking the case \eqreff{kappacon}, the 
 spectral index and the height of the potential have  been calculated, and are
lower than in the original model.

\begin{figure}[t]
\centering
\includegraphics[scale=.8]{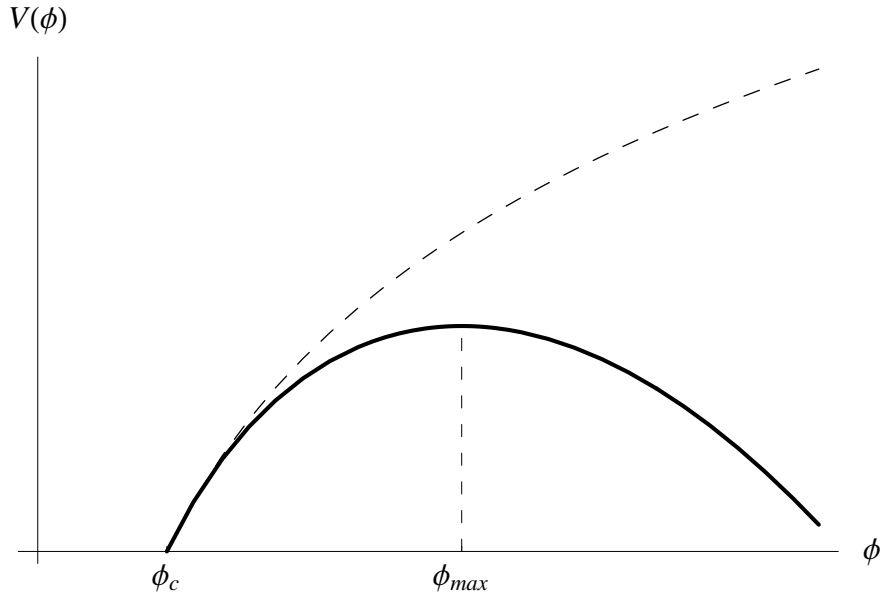}
\caption {Sketch of the inflationary potential for the
$F$/$D$ - term scenario in its simplest form,  without a  tree-level potential
(dashed line) and with a concave-downward  tree-level potential
(continuous line).}
\label{logsketch}
\end{figure}

In the $D$-term case,  $g$ is a gauge coupling which presumably cannot
be small. The vev of the waterfall field has the same cmb normalization
as in the $F$-term case. This vev is expected to be of order the string
scale, relating $D$ term inflation directly to string theory.

There is no $\eta$ problem for the $D$-term model, but the 
tree-level potential $V(\phi)$ is still not expected to be flat
because we are  dealing with a medium-field model where non-renormalizable
terms are out of control.. There is no particular
reason to think that the tree-level $V(\phi)$ 
 will be quadratic, but one may adopt the
quadratic form as a parameterization. The case of positive  mass-squared 
was considered in  \cite{mydterm,km}, and negative mass-squared in
\cite{bl,john}. As in  the $F$ term case, it gives an attractive inflation
model and  with the height of the potential  and  the spectral index both lower
than with the original model.

In both the $F$ and $D$ term models, the inflationary energy scale 
without a tree-level potential is
$V\simeq g^2 \Lambda^4$. Cosmic strings are generically produced with  tension
 $\mu\sim V\half$,  and  the cmb constraint $\mu\half\lsim 10^{15}\GeV$
imposes restrictions on the parameter space.

\section{Tree-level hybrid inflation}

All of the models  considered so far can
give a  spectral index which is consistent with observation
at the time of writing, provided that $N$ is not too far below
the  expected value $\simeq 50$). Now we turn to small- and medium-field
models which at least in their simplest form are ruled out
by their prediction for the spectral index (as always, on the assumption
that the inflaton perturbation generates the curvature perturbation).

Any small- or medium-field model with a concave-upward potential is ruled
out. Such models are of the hybrid type,  unless the potential becomes
concave-downward after cosmological scales leave the horizon.
Taking the fixed point as the origin of symmetries, we
distinguish between
potentials with positive slope as in Figure \ref{smallv3},
and with negative slope as in Figure \ref{smallv4}.

A  negative slope   can arise from
non-perturbative quantum effects \cite{kr}.
More usually, one finds models with positive slope as in Figure
\ref{smallv3},
coming from a tree-level  hybrid inflation model with (say) a quadratic
potential. The
potential including the waterfall field $\chi$ is \cite{andreihybrid}
of the form \eqreff{vord} with $V(\phi)=\frac12 m^2\phi^2$. 

A well-motivated tree-level hybrid inflation model,
called Supernatural Inflation by its authors \cite{rsg},
 uses softly-broken global supersymmetry.
The waterfall field is, in our nomenclature, a light modulus
\cite{bdv}.
In contrast with most models of inflation, the inflationary scale is low
corresponding to  $V_0\quarter\sim M\sub S \sim 10^{10}\GeV$,
the idea being that there is
gravity-mediated supersymmetry breaking both during
inflation and in the vacuum, the only difference in the former case being
that the last term of \eq{vsugra07} has not yet kicked in
  In order to achieve a viable
model the masses $m_\chi$ and $m$ are taken to be respectively
somewhat bigger and smaller than their generic values of order $H_*$.
The observed curvature perturbation
is then obtained with $\lambda'$ just a few orders of magnitude below
1.


The origin $\chi=0$ is taken by the authors to be, in our
nomenclature, a point of enhanced symmetry. The relevant symmetries
cannot be those of the Standard Model because $\vev\chi\sim\mpl$.
After inflation the waterfall field oscillates about its vev, but it
is supposed to decay into SM particles before nucleosynthesis so that
it presents no moduli problem.  This makes the vev another point of
enhanced symmetry, the symmetries now being those of the Standard
Model \cite{bdv}.

As with practically all inflation models, the inflaton
 is invoked just to give inflation and is
not part of any extension of the Standard
Model that has been proposed for other purposes. Models similar in
spirit  have been proposed
(beginning with \cite{bk})
that  are  based on
extensions of the Standard Model that serve other purposes too.
They have an even lower
inflation scale,  corresponding to a mediation strength
stronger than gravitational.
 They  invoke fine tunings, which may however be
reasonable within the context of string theory and branes.
They  can give either ordinary or inverted hybrid inflation, but in both
cases the spectral tilt is practically zero in contradiction with observation.
To avoid this problem though, it seems possible to  
 generate  the curvature perturbation during preheating
 \cite{maretal}.

In considering tree-level hybrid inflation,  one has
to remember that the coupling of the inflaton to the waterfall field
generates a calculable loop correction to the potential,
which can be concave-downward and rescue the model.
 This still leaves a large region of parameter space
in which the one-loop correction from this source is negligible
\cite{myhybrid},
 though in some part of that space one should still
worry about the two-loop correction \cite{rsg}.
In any case
 the coupling of the inflaton to fields other than the waterfall field
can also generate a concave-downward loop correction. We consider this 
possibility next, in the context of the running-mass model. 

A different possibility for generating a concave-downward potential would
be to include the leading non-renormalizable term with a negative sign,
generating a maximum as we discussed already for $F$ and $D$ term inflation.
The possibility has not been investigated at the time of writing.

\section{Running mass models}

The loop correction with  soft supersymmetry breaking generates a running
mass. If the mass belongs to the inflaton we have a 
 running-mass inflation model. The usual model  \cite{ewanrunning} starts 
with the  Supernatural Inflation model that we mentioned earlier.
 At $\phi=\mpl$,
the running mass $m^2(\phi)$ is supposed to be of order $V_0/\mpl^2$,
which is the minimum value in a generic supergravity theory.  The
inflaton is supposed to have couplings (gauge, or maybe Yukawa) that
are not too small, and it is supposed that $m^2(\phi)$ passes through
zero before it stops running. The running associated with a given
loop will stop when $\phi$ falls below the mass of the particle in
the loop.

The potential near $m^2(\phi)=0$ is flat enough to support inflation. To
see this, we can use \eq{vrunning} which is valid over any small range
of $\phi$ and will therefore be valid around the minimum.  It can be
written in the form
\be
V= V_0 \[ 1 + \frac12 \eta_0 \frac{\phi^2}{\mpl^2} \( \ln
\frac\phi{\phi_*} -\frac12 \) \] ,
\ee
which leads to
\be
\mpl \frac{V'}{V_0} = \eta_0 \frac\phi\mpl \ln\frac\phi{\phi_*}
.
\ee
The potential has a maximum or minimum at $\phi=\phi_*$, at which
$\eta=\eta_0$,
 and near which
\be
\eta = \eta_0 \( 1 + \ln\frac{\phi_*}\phi \)
\label{etarun}
.
\ee
A maximum is favoured theoretically, because a minimum requires a
hybrid inflation model with $\phi\sub c$ tuned to be near the minimum.

To estimate $|\eta_0|$, we can make the crude approximation
that
\eq{etarun} is valid at $\phi\sim\mpl$, where $|\eta|$ is supposed to be of
order 1. Then
\be
|\eta_0| \sim 1/\ln(\mpl/\phi_*)
\,.
\ee
This will give $|\eta_0|\ll 1$ if $\phi_*$ is exponentially below
$\mpl$, and with the reasonable requirement $\phi_*\gsim 100\GeV$ it
gives something like $|\eta_0|\sim 10^{-1}$. For a generic value of
$\phi(N)$ this corresponds to $|n-1|\sim 0.1$ which is outside the
observational bound. One can satisfy current observation by choosing
the parameters so that $\phi(N)=\phi_*$ around the middle of the
cosmological range of scales, corresponding to the spectrum having a
maximum at that point \cite{ourrunning}.
 The running of the spectral index at that
point is $d n/d\ln k \simeq - 2\eta_0^2$, and we are requiring
$|\eta_0|\sim 10^{-1}$. This is allowed by
present observations with , though it will soon
be ruled out or confirmed.

To see whether the condition $\phi(N)\simeq \phi_*$ is reasonable, as
well as to calculate the cmb normalization, we need
\be
N(\phi) = -\frac1{|\eta_0|} \ln \( \ln \frac{\phi\sub{end}}{\phi_*}
\ln \frac{\phi_*}\phi \)
\label{Nrun}
.
\ee
If slow-roll inflation ends at $|\eta|\sim 1$, and \eq{etarun} is still
roughly
valid there, $|\eta_0|\ln(\phi_*/\phi\sub{end}) \sim 1$
and \eq{Nrun} requires
roughly $|\eta_0|\simeq \exp(-N|\eta_0|)$ which is more or less compatible
with
$|\eta_0|\sim 0.1$, and also more or less satisfies the cmb normalization
with $V_0\quarter\sim 10^{-10}\GeV$.

A running mass has also been considered in the context of a
two-field modular inflation model \cite{ks,ks2}.
 The  two real fields are components of
a complex field $\Phi$.
The maximum of the
tree-level potential, chosen as $\Phi=0$, represents a point of
enhanced symmetry, and its height is
 $V_0\quarter\sim 10^{10}\GeV$ corresponding to gravity-mediated supersymmetry
breaking.
Writing  $\Phi\equiv|\Phi|e^{i\theta}$, the
potential depends on both $\theta$ and $|\Phi|$.
  The tree-level negative mass-squared defined at
the origin is supposed to have the generic value corresponding to
$|\eta_0|\sim 1$, but interactions cause the mass to run. This turns
the maximum into a crater, and it makes the potential very flat at the
rim so that inflation can take place there.

There is a family of trajectories characterized by the initial value
of $\theta$. The curvature perturbation in this two-field model
was  calculated from the $\delta N$ formalism.
  Near a special value of
$\theta$, chosen as zero,   $\theta$ can be chosen
to reproduce the  cmb normalization is reproduced with
$V_0\quarter\sim M\sub S
\sim 10^{10}\GeV$. It seems to be possible
to reproduce the observed spectral index by choice of parameters.

\section{Large- field models}

Now we turn to large-field models.
They give a significant tensor perturbation $r\sim 10^{-2}$,
 which will be observed or ruled
out in the near future.

 The field variation  cannot actually be extremely large, because
\eq{nofk} requires $\Delta\phi/\mpl < \sqrt{2\epsilon\sub{max}} N
\ll 50$. Two kinds of potential have been considered. One \cite{chaotic}
 is
the Chaotic Inflation
potential  $V\propto \phi^p$ with $p$ an even integer.
The slow-roll parameters  are
\be
\epsilon = \frac{p^2}{2} \, \frac{\mpl^2}{\phi^2}, \qquad
\eta = p(p - 1) \, \frac{\mpl^2}{\phi^2}.
\ee
\noindent Inflation ends
at $\phi_{{\rm end}} \simeq p \mpl$
When cosmological scales leave the
horizon, we find from \eq{nofk} that $\phi= \sqrt{2Np}\, \mpl$, giving
\be
n-1  =   -\frac{2+p}{2N}= -\frac{2+p}{100}, \qquad\qquad
r  =  \frac{4p}{N} = 0.08p
.
\ee

Current observational constraints practically rule out the case $p \geq 4$.
Future observation will rule out or support the remaining case $p= 2$.
The cmb normalization for $V=\frac12 m^2\phi^2$ is $m=1.8\times
10^{13}\GeV$, and for $V=\frac14\lambda \phi^4$ it is $\lambda =7\times
10^{-14}$.  If the curvature perturbation is not generated by the
inflaton, these become upper bounds, and there is no spectral index
constraint.

Another simple possibility    is to use a sinusoidal potential
\be
V= \frac12 V_0 \left[ 1 + \cos\left(\sqrt{2|\eta_0|} \phi/\mpl \right)
\right]
\label{naturalv}
.
\ee
Here, the origin has been taken to be the maximum of the potential,
and $\eta_0<0$ is the value of $\eta$ there.  This was called Natural
Inflation by its authors \cite{natural}. The vev is at $\vev\phi=-\pi
\mpl/\sqrt{2|\eta_0|}$.

 With this potential $\phi(N)$
is given by
\be
\sin\( \sqrt{|\eta_0|\over 2} {\phi\over\mpl}\)
= \sqrt{1 \over 1 + |\eta_0|} \,e^{-N |\eta_0|}
\label{phinat}
,
\ee
leading to
\be
\epsilon =  \frac1{2N} \frac{2N|\eta_0|}{e^{2N|\eta_0|}-1}, \qquad\qquad
\eta =  \epsilon -  |\eta_0|
\label{naturaleps} .
\ee
The maximum is at $\phi=0$, and eternal inflation can take place there
providing the initial condition for observable inflation. 
But if $N|\eta_0|\ll  1$,  observable inflation itself will not
begin until the potential is near the minimum, 
corresponds to the `chaotic inflation' potential $V=\frac12
m^2\phi^2$.  
 The prediction in the $r$-$n$ plane is shown in Figures \ref{rvsn} and 
\ref{newrvsn}. 
We see that the current bound on $n$ requires $r\gsim 10^{-2}$. 
This  means that Natural Inflation will eventually be confirmed or ruled out,
 though it may turn out to be indistinguishable from chaotic inflation.

Large-field models are difficult to
understand within the
generally accepted rules for constructing field theories
beyond the Standard Model, whereby the higher order terms in the expansion
\eqreff{power} are under  control only for $\phi\ll \mpl$.
Some possibilities do exist though.

First, the inflationary trajectory may lie in the
space of many fields, corresponding say  $\phi=
\sum_{i=1}^N  a_i \phi_i/\sqrt{\sum a_i^2} $. Then, with say all $a_i$ equal,
we can have $\phi\gg \mpl$ with each  $\phi_i\ll\mpl$. This was called
Assisted Inflation by its authors \cite{assisted}.
 At first
sight one might think that the proposal lacks content, since a rotation
of the field basis can always make $\phi$ one of the fields.
The point though is that the field theory may select a particular basis,
as the one in which  the power series \eqreff{power} is expected  to be
relevant.  It has been argued \cite{nflation} that this
will be the case if each $\phi_i$ has a  sinusoidal potential,
leading to what they
called $N$-flation. Then, if inflation takes place near the minimum
of the potential one can have $\phi^2$ chaotic inflation even though
the proportionality $V\propto \phi^2$  does not persist up to the
Planck scale.

A second possibility is for  the inflationary trajectory may
wind many times around the fixed point of the symmetries, at a distance
$\lsim \mpl$ from that point. Something like this has been suggested in
the context of string theory \cite{knp},
giving
 a sinusoidal  potential corresponding to Natural Inflation. Finally,
it may be possible  to evade the general
rule that \eq{power} is out of control at $\phi\gg\mpl$,
 if the field theory is derived from a special higher-dimensional setup.
This is the idea of Gauge Inflation \cite{extranat,kw,pst},
where the inflaton is the fifth component
of a gauge field  living in a 5-d  theory, which becomes  a PNGB  in the
4-d theory. This again can give a sinusoidal potential.
None of these
proposals allows $V$ to increase continually up to the Planck scale, in
the spirit of the Chaotic Inflation proposal.

\section{Warm Inflation}

In all of the inflation models mentioned so far, energy loss by the
inflaton field $\phi$
is assumed to be negligible on the grounds that $\phi$
changes only slowly with time. Including this energy loss will give an
equation of the form
\be
\ddot\phi + (3H + \Gamma) \dot \phi + V' = 0
,
\ee
where $\Gamma$ is some time-dependent quantity. The warm inflation
model \cite{warm}
 assumes that $\Gamma$ is significant, or even dominant ($\Gamma
\gg H$).

The extent to which warm inflation is possible was investigated in the
GUT hybrid inflation model \cite{warm3}
 using an earlier  calculation of the energy loss \cite{warm2}.
 It does not occur in the original GUT hybrid model but apparently
can occur if the
inflaton has  a suitable interaction with a spin-half particle.
The curvature perturbation in warm inflation receives a contribution
from the thermal fluctuation, which dominates the contribution of the
vacuum fluctuation if $\Gamma$ is dominant.

\section{Present status and outlook}

    \begin{figure*}

\centering\includegraphics[angle=270,width=1.0\linewidth]
{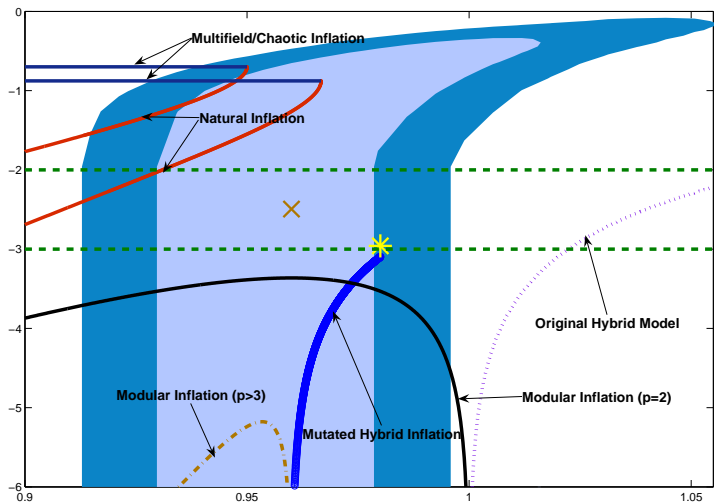}
        \caption{The shaded regions are the allowed by observation as in
Figure \ref{rvsn}, and the predictions are described in the text. Planned
observation will detect $r$ or give a limit $r<10^{-2}$, and $r<10^{-3}$
will probably never be observed.}
                \label{newrvsn}
    \end{figure*}

Figure \ref{newrvsn} summarizes most of the  predictions that we have been
discussing,   always assuming that the inflaton
perturbation generates the curvature perturbation.
(Recall that the alternative was considered in Section
\ref{sgenpred}.)

Consider first
small- and medium-field models. 
For these models the tilt is directly
related to the curvature of the potential,  $n-1=2\eta$. As a result,
the recently-observation negative tilt has had a dramatic effect,
ruling out  whole classes of otherwise attractive models.
These include the original   tree-level
hybrid inflation model, in particular  those rather well-motivated
versions 
which invoke during inflation the  vacuum  supersymmetry-breaking
mechanism. The running-mass variant of tree-level
hybrid inflation is  not yet ruled out, but it will be if
 the  observational  bound on the running of $n$
gets much tighter.

Among simple single-field slow-roll models, the ones that agree with
observation are
modular inflation,  and hybrid inflation with a concave-downward potential.
The latter can be
achieved by what are usually termed simply $F$- and $D$-term inflation,
involving the loop correction generated by spontaneously broken global
supersymmetry. They can also be achieved by mutated hybrid inflation.

All of these simple  models give (exactly or as what should be a reasonable
 approximation)
a distinctive prediction for the
scale-dependence of the tilt, of the form
\be
n-1 =  - \( \frac{p-1}{p-2} \) \frac 2{N(k)}
.
\ee
This gives the scale-dependence (running)
\be
\frac12 \frac{dn}{d\ln k} = -\( \frac{p-2}{p-1} \) \( \frac{n-1}{2} \)^2
.
\ee
 Several years down the line it might be possible to measure this level of
 running,
for instance through a measurement of the 21-cm anisotropy. A confirmation
of the above prediction would select within observational uncertainty
values for both $N$ and $p$. If the former were in the relatively
narrow range compatible with post-inflationary cosmology,
one would probably be convinced that
 that a model with the relevant $p$  is correct. That would be a truly
remarkable development, since it would imply a high
inflation scale  $V\quarter \sim 10^{15}\GeV$ and with a sufficiently
accurate value of $N$ the reheat temperature would also be determined
(assuming  continuous radiation domination after inflation).

Now consider  the large-field models.
The  prediction for $r$ and $n$ is compatible with
observation for $V\propto\phi^2$,  and for Natural Inflation if
the period of the potential is not too small.
 {}From Figure \ref{rvsn} it is clear that a joint
measurement of $r$ and $n$ can rule out these models. Conversely,
a measurement of $r$ and $n$ in agreement with one of them would be
very suggestive.
Again, many years down the line further confirmation could come from
a measurement of the running of $n(k)$ and $r(k)$, which goes along the
lines indicated in Figure \ref{rvsn}. And, again, if such a measurement
were compatible with a sensible value for $N$ one would be convinced
about the validity of the model, implying again the high inflation scale
now $V\quarter\sim 10^{16}\GeV$.

\section{Acknowledgements}
I thank my collaborators, and in particular Lotfi Boubekeur and Laila
Alabidi who supplied  the Figures. The research is supported by
 PPARC grant
PPA/G/S/2003/00076. DHL is  supported by
PP/D000394/1   and by EU grants
MRTN-CT-2004-503369 and MRTN-CT-2006-035863.



\printindex
\end{document}